\documentclass[a4,12pt,twocolumn]{article}
\usepackage{amsmath}
\usepackage{amssymb}
\usepackage{url}
\usepackage{graphicx}
\usepackage{dcolumn}
\usepackage{bm}
\usepackage[caption=false]{subfig}
\usepackage{xcolor}
\usepackage[numbers,sort&compress]{natbib}
\usepackage{amssymb}
\usepackage{hyperref}
\hypersetup{
    colorlinks=true,
    linkcolor=blue,
    filecolor=magenta, 
    citecolor=blue,
    urlcolor=cyan,
    pdfpagemode=FullScreen,
}
\usepackage{soul}
\usepackage{comment}
\usepackage[mathlines]{lineno}
\usepackage{soul}

\graphicspath{{./FIG-dynSpiral/}}

\begin{document}
\title{Physical Principles of Size and Frequency Scaling of Active Cytoskeletal Spirals}

\author{Aman Soni$^\dag$, Shivani A. Yadav$^\dag$ and Chaitanya A. Athale\footnote{All correspondence to be directed to cathale@iiserpune.ac.in},\\
Div. of Biology, IISER Pune, Dr. Homi Bhabha Road,\\ Pashan, Pune 411008, India.\\
$^\dag$ These authors contributed equally to the manuscript.}%



\date{}
\onecolumn
\maketitle

\section*{Abstract}

Cytoskeletal filaments transported by surface immobilized molecular motors with one end pinned to the surface have been observed to spiral in a myosin-driven actin `gliding assay'. The radius of the spiral was shown to scale with motor density with an exponent of -1/3, while the frequency was theoretically predicted to scale with an exponent of 4/3.  While both the spiraling radius and frequency depend on motor density, the theory assumed independence of filament length, and remained to be tested on cytoskeletal systems other than actin-myosin. Here, we reconstitute dynein-driven microtubule spiraling and compare experiments to theory and numerical simulations. We characterize the scaling laws of spiraling MTs and find the radius dependence on force density to be consistent with previous results. Frequency on the other hand scales with force density with an exponent of $\sim 1/3$, contrary to previous predictions. 
We also predict that the spiral radius scales proportionally and the frequency scales inversely with filament length, both with an exponent of $\sim1/3$.
A model of variable persistence length best explains the length dependence observed in experiments. Our findings that reconcile theory, simulations, and experiments improve our understanding of the role of cytoskeletal filament elasticity, mechanics of microtubule buckling and motor transport and the physical principles of active filaments.

\section*{Keywords} Microtubule, persistence length, motor, spiral, mechanics, dynein, scaling theory.
\section*{Introduction}
	Microtubules (MTs) are cytoskeletal polymers composed of $\alpha$- and $\beta$-tubulin heterodimers and essential for eukaryotic cell physiology \cite{desai1997microtubule}. Filaments of MTs exhibit a polarity based on subunit geometry \cite{bergen-borisy1980} that correlates with kinetic polarity with plus-ends growing on average faster than minus-ends. 
Motor proteins associated with MTs walk either to the plus-ends, kinesins or the minus-ends, dyneins, with some variations \cite{walker1990drosophila,mcdonald1990kinesin}. MTs and motors combine to play a key roles in sub-cellular force generation \cite{Sanghavi2023}, transport \cite{schroer1988role,schnapp1989dynein} and positioning of organelles \cite{schatten1982taxol}, as well as spindle assembly and segregation \cite{Walczak:1998tp} and sperm motility \cite{gray1955movement}. How the mechanics of motors and microtubules result in cellular function remains an active area of investigation.

Optical tweezer based single-molecule force-spectroscopy has allowed the measurement of force generation at the individual molecule level \cite{block1990bead,reckpeterson2006}. However, inside cells both MTs and motor proteins work in collectives of many motors walking in complex environment, suggesting that understanding collective mechanics is important to better understand their role in cells. A conventional approach taken to examine collective effects of motor transport is the {\it in vitro} reconstitution of surface-immobilized motors that then drive the movement of MTs in presence of ATP, i.e. the `gliding assay'. 
ATP-dependent stepping of motor heads anchored by their tails results in filament translocation in the direction opposite to the polarity of the motors \cite{howard2001mechanics}. 
Such assays are common to both MT-motor and actin-myosin systems and when an end of such a filament is either clamped or pinned it results in spiraling or flapping, respectively. Pinning sites have been reported to arise from either surface defects \cite{maloney2011effects} or inactive motor fraction \cite{bourdieu1995}, and the spiraling patterns of actin \cite{bourdieu1995} and bending of clamped MTs \cite{gittes1996directional} have been analyzed to derive relations between motor and filament mechanics as well as estimate mechanical properties of cytoskeletal filaments. More recently, we reported the emergence of wave-like oscillatory patterns of single microtubule filaments clamped at one end driven by dynein motors in a `gliding assay' \cite{Yadav2024}. Our numerical simulations could predict experimentally observed patterns of filament dynamics transitioning from flapping, through oscillations to looping, in a manner that dependend on force generator (motor) density and MT length. Thus, theory and experiments of such spatial patterns have been widely used to examine buckling instabilities and mechanics of both motors and filaments. A general theory of such such tip-constrained gliding assays was developed predicting the scaling of spiral radius to force generated by molecular motors, that showed agreement with experiments with myosin driven actin spirals \cite{bourdieu1995}. The scaling argument predicts that the radius (R) will scale with linear force density ($f$) as $R \sim (k_BT\ell_{p}/f)^{1/3}$, i.e. the spiraling radius is expected to scale with the inverse cube root of force density. 
Similarly, a theoretical scaling exponent of 4/3 was predicted for the spiraling frequency with force density was shown in numerical simulation of self-propelled filament \cite{bourdieu1995,fily2020buckling}. While some of the theoretical predictions of spiral radius scaling with force  have been tested, their generality is yet to be established. 
In addition, the spiraling frequency has to our knowledge not been experimentally measured, pointing to a need to test the validity of the theoretical predictions.  

Mechanically, MTs are distinct from the actin filaments, as physical length scale and persistence length are separated by three orders of magnitude \cite{gittes1993flexural}.
Furthermore, the persistence length of microtubules has been measured in multiple studies, resulting in diverse outcomes. Some studies suggest a length-independent persistence length \cite{gittes1993flexural,kawaguchi2008temperature,kikumoto2006flexural,felgner1996flexural}, while others propose a length-dependence \cite{pampaloni2006thermal,kurachi1995buckling,takasone2002flexural}. Traditionally, microtubules have been assumed to be elastically isotropic \cite{gittes1993flexural}. However, recent evidence indicates that microtubules exhibit elastic anisotropy \cite{kis2002nanomechanics}. This anisotropy arises from the differential strengths of molecular interactions within tubulin dimers -- longitudinal bonds are stronger than lateral bonds between adjacent protofilaments \cite{sept2003physical}. The existence of elastic anisotropy-induced length-dependent persistence length in MT could result in length dependence in spiraling radius and frequency, which was not reported earlier.

Here, we have proceeded to reconstitute steady state MT spirals arising from end-pinned filaments driven by sheets of dynein motors in a `gliding assay'. We analyze the spiral size and frequency dependence on motor density and filament length and compare our findings to theory and numerical simulations.


\section*{Model}
\subsection*{Mechanics of microtubules and motors}
The \textit{in silico} model of the gliding assay consists of only two components: (i) microtubules and (ii) molecular motors. While the model is extensively described elsewhere \cite{nedelec2007collective,Nedelec2002}, the key features of the models are highlighted here. 

(i) Microtubules are modeled as a discretized elastic inextensible rods with elasticity modeled based on Euler beam mechanics and the individual points of discretization following bending elasticity. 

(ii) Molecular motors are modeled as spring-like explicit force generators with a linear force-velocity (F-V) relation, expressed as follows:
\begin{equation}
    V(F) = V_{0}(1-F_{||}/F_{stall})
    \label{eq:forcevel}
\end{equation}
where $V_0$ is unloaded motor velocity, $F_{||}$ is the projection of the load force (F) acting on a motor and $F_{stall}$ is the motor stall force. The attachment and detachment to microtubules is modeled stochastically. The detachment rate ($k_{d}$) is modeled to be force-dependent based on Kramer's law \cite{kramers1940brownian}: 
\begin{equation}
    k_{d}(F) = k'_{d}\text{exp}(|F|/F_{d})
\end{equation}
Where the $k'_{d}$ is the basal detachment rate of a motor from an MT, $F$ is the restoring spring force $F=k \delta l$, proportional to stiffness, $k$ and extension $\delta l$ and $F_d$ is the characteristic detachment force. The detachment rate is assumed to be symmetric with respect to direction of extension. The detachment exponentially grows with the increase in the force. As the system of interest operates in a low Reynolds number regime, viscous forces dominate, and inertial effects can be ignored. The simulations are updated using an overdamped Langevin equation with uncorrelated noise incorporating the effect of thermal energy. Filaments are pinned at one end through a pivot with a ten-fold larger stiffness constant 
compared to that of the motors, sufficient to restrict translational mobility, maintaining complete rotational freedom at the pinned end (Figure \ref{fig:simintro}A). 

\subsection*{Models of MTs persistence length}
In this study, we simulate MTs, based on two alternative models of the elastic properties proposed in the literature: (a) constant persistence length ($\ell_{p}^{c}$), based on an isotropic elastic model of filaments and previous reports \cite{gittes1993flexural} and (b) a variable persistence length model ($\ell_{p}^{v}(L)$) that can be explained based on the reported anisotropic material properties \cite{kis2002nanomechanics}. 
We approximate the anisotropic elasticity of MTs by 
an effective variable persistence length model that changes with filament length as reported previously \cite{pampaloni2006thermal} with:  
\begin{equation}
 \ell_{p}^{v} (L) = \ell_{p}^{\infty}(1+ 3EI/GAkL^{2})^{-1} 
\label{eq:pampaloniK}
\end{equation}
where $l_{p}^{\infty}$ = 6.3$\times 10^ 3$ $\mu$m and $3EI/GAk$ = 441 $\mu$m$^2$ as parameterized earlier. While this equation suggests that $\ell_{p}^{v}(L)$ reaches an asymptotic value at high values of length, $L > 100$ $\mu$m, these lengths are unphysiological. However at physiological length scales of $\sim$ 1 to 10 $\mu$m, $\ell_{p}^{v}(L)$ changes by three orders of magnitude in length scale (Figure \ref{fig:persistence-length-plots}A), in contrast to the constant persistence length model (Figure \ref{fig:persistence-length-plots}B). 


\section*{Results}
\subsection*{Emergence of MT spirals from pinned filaments driven by immobilized molecular motors}
\begin{figure}[ht!]
    \includegraphics[width=1\linewidth]{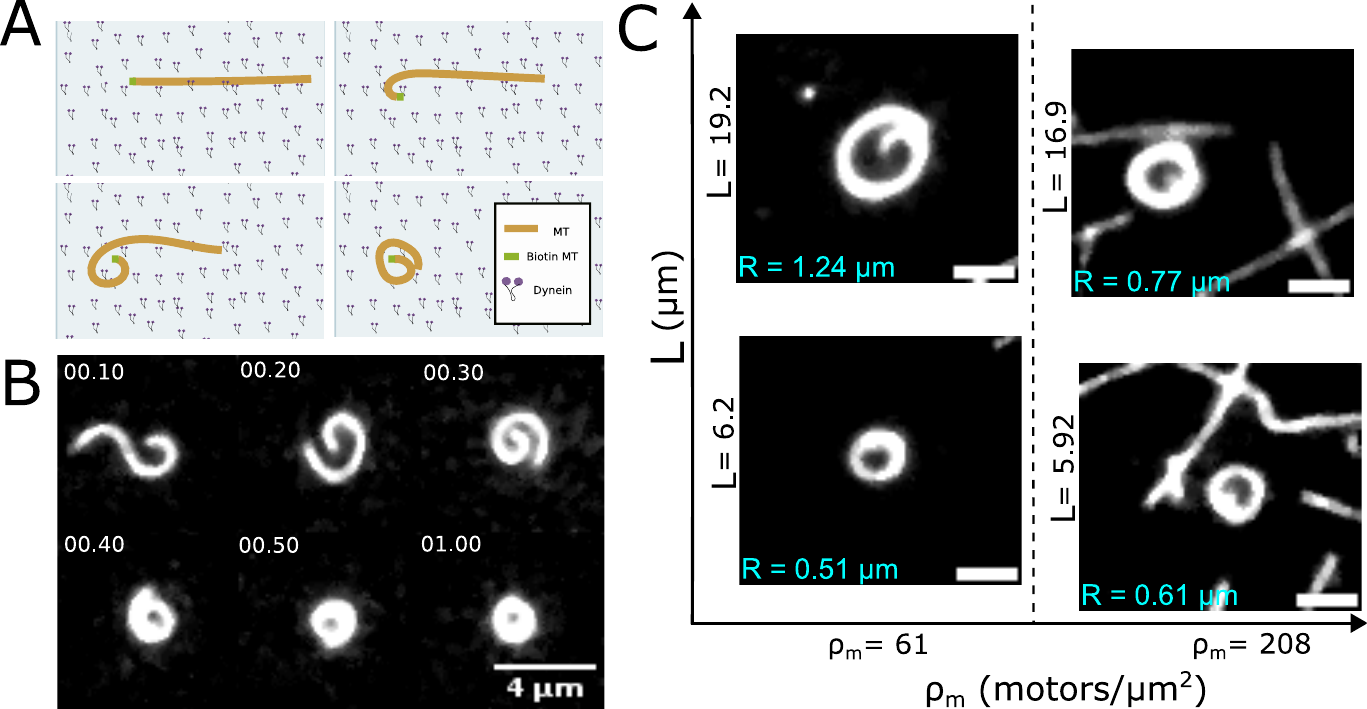}
    \caption{\textbf{Dynein driven MT spirals reconstituted \textit{in vitro}.} (A) Schematic representing formation of dynein driven MT spiral in a modified gliding assay where the leading end of the MT is pinned to the cover-slip surface using biotin-streptavidin chemistry. (B) Montage of a representative MT spiral being formed as a result of pinning of MT leading end and stepping of dynein in presence of ATP where $L$ = 6.8 $\mu$m and dynein density was 61 motors/$\mu$m$^2$. Time: mm:ss. (C) Representative images of steady state MT spirals selected from time-series are displayed for different values of MT length ($L$) and motor density ($\rho_{m}$) showing a length (y-axis) and motor-density (x-axis) dependence of  the measured spiral radius (R). Scalebar: 2 $\mu$m.} 
    \label{fig:experimental-stepup}
\end{figure}

\begin{figure}[ht!]
    \centering
    \includegraphics[width=1\linewidth]{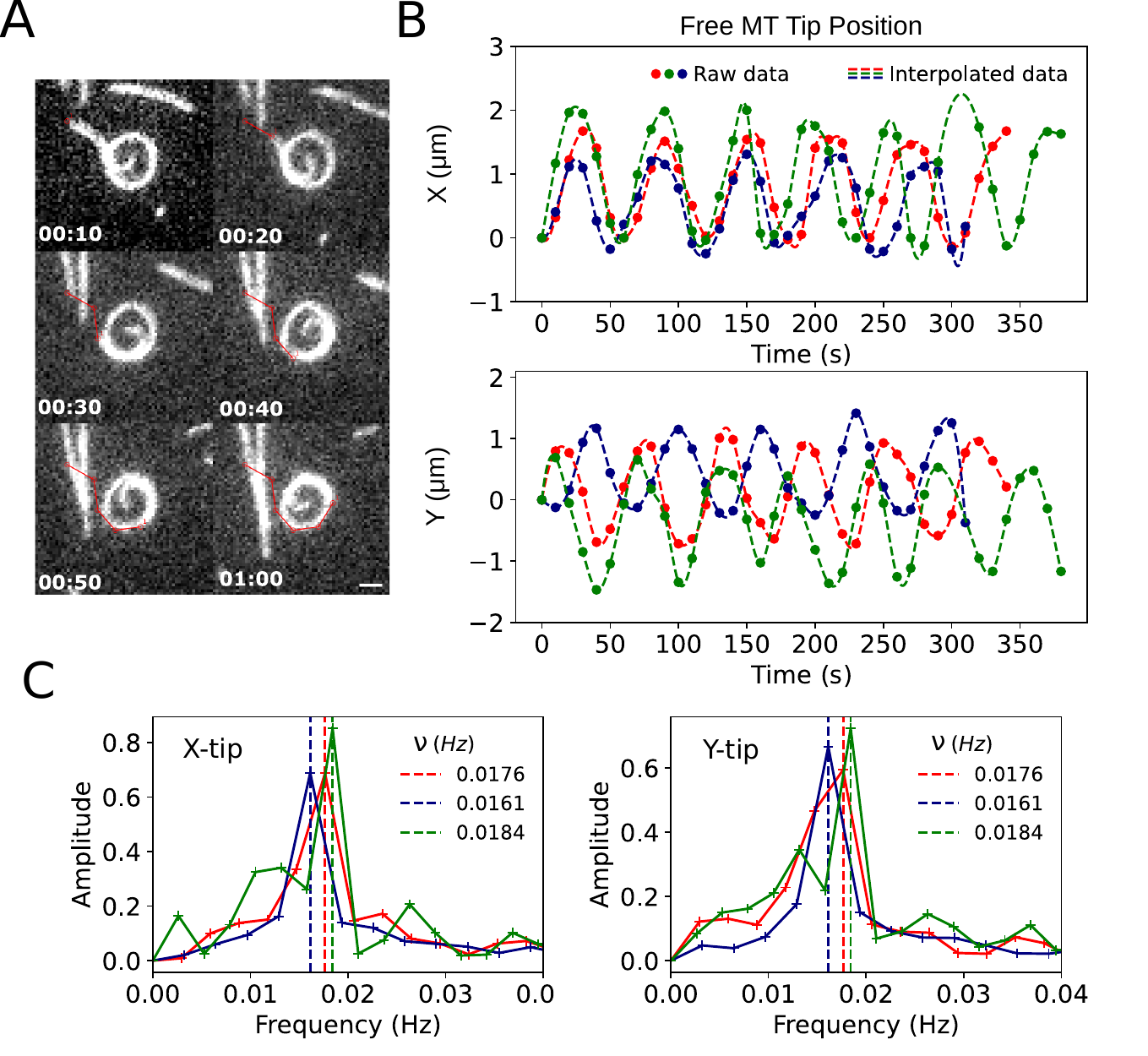}
    \caption{\textbf{Dynamics of MT spiral tips and frequency estimation.} (A) A montage showing tracking of the MT free tip (minus end) of a representative MT as it undergoes dynein driven spiraling. The X and Y position with time of the free MT tip was tracked interactively using MtrackJ (see materials and methods). Scalebar: 2 $\mu$m.  Time: mm:ss. (B) The {\it (top)} x- and {\it (bottom)} y-coordinates of the free-tip of 3 representative spiraling MTs is plotted as a function of time (\textbullet, colors: individual MT tips). The data was smoothed using cubic interpolation and is overlaid (- - -). (C) The amplitude (y-axis) is plotted with the frequency (x-axis) from fast Fourier transform (FFT) of the smoothed data. The dominant frequency 
($\nu$) is marked for each time-series as a vertical line (- - -). Colors: individual MTs. Legend: $\nu$ values.} 
    \label{fig:freq_track}
\end{figure}

We have reconstituted dynein-driven MT spirals in a modified gliding assay setup where the leading MT tip is pinned to the surface. A minimal yeast cytoplasmic dynein construct with GFP on its tail was used as the force generator. The motor is anchored to the glass through an anti-GFP nanobody, that maintain its orientation with the heads free to engage with MTs. The plus-end of MT is pinned to the surface by using biotin-streptavidin chemistry 
(Figure \ref{fig:experimental-stepup}A) similar to that described by us in a previous study \citep{Yadav2024} and elaborated on in the Methods section. 
We observe MTs that are pinned by a very small segment of the plus-end appear to buckle and begin to rotate around the pinning point, eventually forming a steady state spiral 
(Figure \ref{fig:experimental-stepup}B, Video \ref{vid:spiral_expt}). In representative data of two densities and two MT lengths, we find the spiral radius could be affected by dynein motor density 
and MT length (Figure \ref{fig:experimental-stepup}C). We interactively track the dynamics of the free, mobile tip of the MT (Figure \ref{fig:freq_track}A) and use these to quantify the position-time oscillations of the X- and Y-coordinates of the free ends, that appear comparable for multiple spirals (Figure \ref{fig:freq_track}B). We observe a dominant frequency of 0.016 to 0.018 Hz (16-18 mHz) from fast Fourier transform analysis of the position-time oscillatory dynamics 
(Figure \ref{fig:freq_track}C). This frequency is comparable to the wave-like oscillation frequency, as described in a previous study \citep{Yadav2024}. This suggests that the rate of oscillations are robust and comparable, since the mechanics of driving motors (dynein) and MT mechanics are conserved in both these systems.

Based on these observations, we proceeded to develop both theory and numerical simulation to examine general principles of pinned filaments driven by forces generated by molecular motors. 

\subsection*{Numerical simulation with minimal model components reproduces spiraling patterns}
\begin{figure}
    \centering
    \includegraphics[width=1\linewidth]{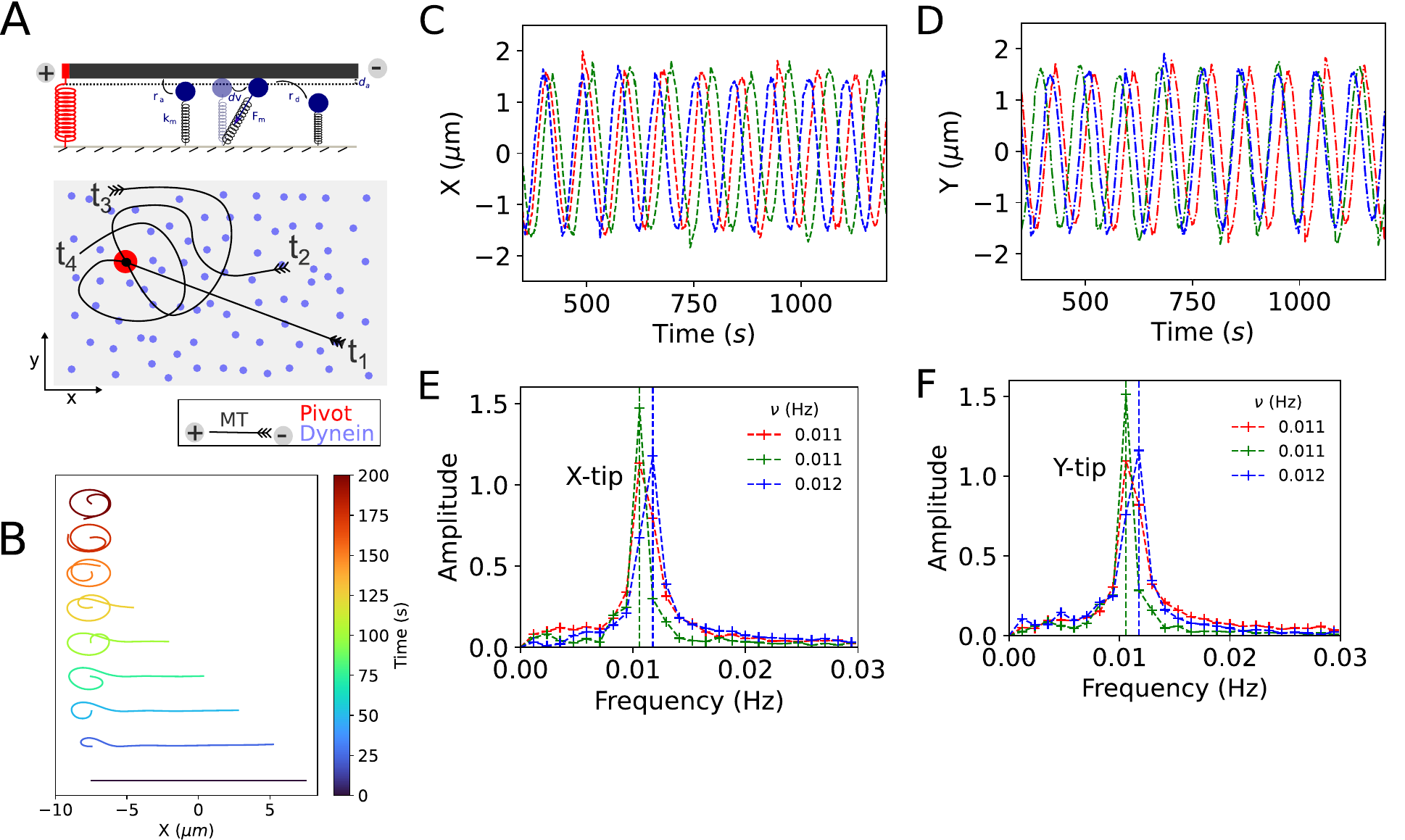}
    \caption{
    \textbf{Simulating motor-driven microtubule spiraling in 2D.} (A) The schematic represents a single microtubule (gray) pinned at the plus end  by a pivot (red) and interacting with cytoplasmic dynein motors (blue) uniformly distributed in 2D space. The force generated by cytoplasmic dyneins bound to the pinned MT results in bending and buckling. The pivot (red) is a spring with a high stiffness constant that allows rotational motion. (B) The contours of one representative MT pinned and driven to form spirals by motor forces are projected in time along the y-axis. The time interval between contours is 25 seconds. Colorbar: time in seconds. (C,D) Temporal evolution of the free microtubule (MT) tip in the (C) y- and (D) x-coordinates, plotted after they have attained steady-state (spiral) configuration. 
 (E,F) The amplitude as a function of frequency of three representative (E) x- and (F) y-coordinates of MT tip positions (colors: individual filaments) is plotted using fast Fourier transform (FFT), as described in the Methods section. Vertical dashed lines: dominant frequency. Simulation parameters: $L$ = 15 $\mu$m, $\rho_{m}$ = 100 motors/$\mu$m$^{2}$, and total simulation time = 1200s.} 
    \label{fig:simintro}
\end{figure}

We performed simulations of spiraling MT based on the description of the numerical model with molecular motor explicitly modeled and semi-flexible polymer treatment of MT. MT was initialized as being parallel to the long axis of the simulation cell and the force exerted by motors acts along the length (Figure \ref{fig:simintro}A). The restriction on translation at the pinned end causes the filament to buckle at sufficiently high forces, while the rotational degrees of freedom result in curved profiles that spiral and converge eventually into a steady state limiting circle ($\sim$200 s)  
(Figure \ref{fig:simintro}B). We plot the position of the microtubule's free tip after attaining a steady-state spiraling configuration persisting through the remainder of the simulation (up to 1200 seconds) (Figure \ref{fig:simintro}C,D). To quantify the oscillatory behavior, we analyzed the steady-state trajectory of the free tip and extracted the dominant frequency using a fast Fourier transform (FFT) applied to the trajectory data from 350 seconds onward (the same trajectory length is used consistently throughout the analysis).
Both the X and Y coordinates showed a frequency of approximately $\sim$0.01-0.02 Hz. Since the filament undergoes circular motion, the X and Y dynamics are coupled and exhibit identical frequencies. This simulated frequency is in close agreement with experimentally observed values (Figure \ref{fig:simintro}E,F). To analyze the spiraling phenomenon, we develop scaling arguments in the following sections and compare them with numerical simulations and experimental results.

\subsection*{Theoretical scaling relationship of spiraling radius with motor density and MT Length}
In order to theoretically explore the scaling of the spiraling radius in motor-driven MTs, we derive a simple asymptotic scaling argument by balancing the bending energy and work done by motor proteins. The work done by motors proteins along the contour for spiraling MTs is $W = \int_{0}^{2\pi} \int_{0}^{L} f dl \cdot dq$, where $f dl$ represents the combination of the force density $f$ applied over a differential length element $dl$ along the MT contour. This results in the displacement of the $dl$ element by an infinitesimal length $dq$ along the spiral. By approximating motor-driven displacement of the filament along the limiting circle, at steady state, as $dq = R d\theta$, the work done then scales as $W \sim f R  L$.  Then we proceeded to estimate bending energy on the same intervals, $E_{b} = \int_{0}^{2\pi} d\theta \int_{0}^{L} \kappa_{r}\lambda^{2}(s) \text{ } ds$, where $\lambda(s)$ represents the local curvature and $\kappa_r$ is flexural rigidity.  Considering the spiral to asymptotically approach a limiting circle, we approximate the local curvature of the spiral as  $\lambda(s) = 1/R$,  where $R$ is the radius of the circle and $s$ is the arc length. This leads us to arrive at a relation for bending energy $E_{b} \sim  \kappa_{r}L/R^{2}$.
Finally, equating the bending energy with the work done, we arrive at a scaling relation for the radius of the circle: 
\begin{equation}
R \sim (\ell_{p}k_{B}T)/f)^{1/3}
\label{eqn:new-scaling-argument}
\end{equation}
where we have substituted for the flexural rigidity by $\kappa_{r} \equiv k_{B}T\ell_{p}$ 
Similar scaling arguments have also been previously made using force and torque balance for self-propelled polymers to arrive at a similar scaling relation for spiral radius $R \sim f^{-1/3}$ \cite{chelakkot2012flow,bourdieu1995,chelakkot2014flagellar}. However, unlike previous studies, 
we do not assume a constant persistence length; instead, we test the hypothesis of the filament length dependence of persistence length by 
quantifying the scaling 
of spiral radius with motor density and filament length.


\subsection*{Reconciling spiral radius scaling predictions with motor density from experiments and simulations}

\begin{figure}
    \centering
    \includegraphics[width=1\linewidth]{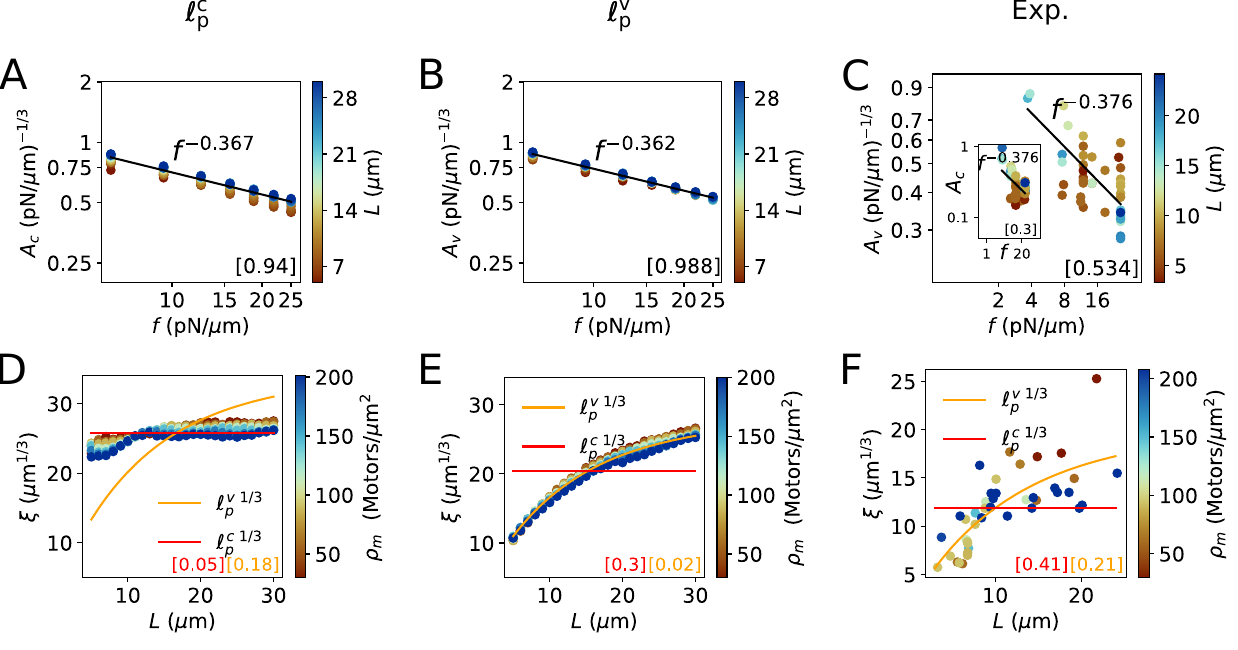}
    \caption{\textbf{Scaling of spiraling radius as a function of motor density and MT length} Rescaled spiraling radius, defined as either $A_c \equiv R / (k_B T \ell_p^c)^{1/3}$ (for constant persistence length) or $A_v \equiv R / (k_B T \ell_p^v)^{1/3}$ (for variable persistence length), is plotted as a function of force density ($f$), with microtubule (MT) length color-coded. (A-B) Simulation performed using (A) constant persistence length (B) with persistence length. (C) Rescaled experimental data using variable persistence length (inset) shows rescaling using constant persistence length. A power-law of the form $A_{c/v} \sim f^{\alpha}$ is fitted to the data (solid black line, ---), with the corresponding $R^2$ value displayed in the lower left corner.  (D–F) Rescaled spiraling radius $\xi \equiv R(f/k_B T)^{1/3}$ of spiraling MTs is plotted as a function of MT length, with motor density color-coded. Scaling relations of the form $\xi \sim \ell_p^{c~1/3}$ and $\xi \sim \ell_p^{v~1/3}$ are fitted to the data and shown as red and orange lines, respectively, with RMSRE (see method section) indicated in the lower right corner. Simulations were performed across MT lengths $L$ from 5 to 30 $\mu$m and motor densities $\rho_m$ from 50 to 200 motors/$\mu$m$^{2}$, with $\ell_{p}^{c}$ set to 5 mm for constant persistence length (A,D) and $\ell_{p}^{v}(L) = 6.3/(1 + 441/L^2)$ mm for variable persistence length (B,E). Simulated data is represented by a mean of 20 repetitions.}
    \label{fig:radius-scaling-scaled}
\end{figure}

To identify a scaling relation, we achieved data collapse by appropriately rescaling the  variables (Figure \ref{fig:raw-radius}A-C). The rescaled radii, $A_c$ and $A_v$, were obtained by normalizing the spiraling radius ($R$) using the factors $(\ell_p^{c} k_B T)^{1/3}$ and $(\ell_p^{v}k_B T)^{1/3}$ for constant and variable persistence lengths 
respectively, to examine the dependence of radius on force density alone, by eliminating the influence of other parameters. As the rescaled data formed a tight cluster of data points, we fit a power law of the form $A_{c/v}$  
and obtained scaling exponents $\alpha = -0.36$ for both constant and variable persistence length simulations (Figure~\ref{fig:radius-scaling-scaled}A,B).
The values of the exponent are very similar to the theoretically predicted exponent and with previously reported results \cite{bourdieu1995,fily2020buckling}. 
We find our experimentally measured scaling relation of the radius is consistent with both constant and variable persistence length simulations (Figure \ref{fig:radius-scaling-scaled}C). Since the persistence length is fixed for a given filament length—regardless of whether a constant or length-dependent model is used—the scaling exponent as a function of force density remains unchanged across models (Figures \ref{fig:radius-scaling-scaled}A,B;\ref{fig:effect-of-trap-stiffness-radius};\ref{fig:scaling-stiffness-exponent}A,B).

However, we observe a slight deviation in the scaling of experimental ($\alpha \approx -0.37$) and simulation data ($\alpha \approx -0.36$) compared to the theoretical prediction ($\alpha = -0.33$). This discrepancy can be explained by the choice of pivot stiffness which was used to match the experimental observations. When simulation performed with varying stiffness of pivot scaling exponent approaches theoretical prediction of -1/3 (Figure \ref{fig:effect-of-trap-stiffness-radius};\ref{fig:scaling-stiffness-exponent}A,B). Traps with lower stiffness impose limited constraints on translational motion compared to those with higher stiffness. Consequently, curvature relaxation tends to occur through translational motion in the vicinity of the pinned tip, rather than being directed into spiraling motion. This localized dissipation of bending at pinned may lead to deviations from the scaling behavior predicted by theory.

\subsection*{Variable persistence length model can explain spiral radius scaling with MT length}
To quantify the scaling of the spiraling radius with MT length (Figure~\ref{fig:raw-radius}D-F), we decouple the effect of force density by rearranging the terms in the Equation \ref{eqn:new-scaling-argument} and defining rescaled spiraling radius $ \xi \equiv R (f/k_BT)^{1/3}$.  In the relation of radius scaling with linear force density (Equation \ref{eqn:new-scaling-argument}), there is no explicit length dependence. However, there could be implicit length dependence via persistence length due elastic anisotropy and can lead to non-trivial scaling behavior. We therefore predict that the rescaled radius scales with microtubule length proportionally to the persistence length, in the following manner:

\begin{equation}
\xi(L) \sim \ell_p^{1/3}
\end{equation}

To test the prediction of this scaling argument, we fit simulation data--generated using either a constant or length-dependent persistence length--to both $\ell^{c~1/3}$ and $\ell_p^{v~1/3}$ scaling functions. The generality and validity of the scaling argument holds only when the fitted scaling function reflects the input persistence length used in the simulation.
In simulations with constant persistence length, $\xi$ is better explained by the scaling $\sim \ell_p^{c~1/3}$ (root mean square of relative error, RMSRE: 0.05 $<$ 0.18; Figure~\ref{fig:radius-scaling-scaled}D). Similarly, simulations with variable persistence length, $\xi$ is better captured by the scaling $\sim \ell_p^{v~1/3}$ (RMSRE: 0.02 $<$ 0.3; Figure~\ref{fig:radius-scaling-scaled}E). These results indicate that the proposed scaling argument holds true for either constant or variable persistence length models. Thus, the scaling relation can be used to distinguish the nature of persistence length. 
Experimental data were fit by both models, and found to be better described by the variable persistence length model (RMSRE: 0.21$<$ 0.41), indicating that microtubule persistence length is not constant but instead varies with filament length  (Figure~\ref{fig:radius-scaling-scaled}F). 

Based on these observations of scaling in the spatial domain, we proceeded to examine the theory of time domain scaling of filament spiraling and reconcile it with experiments.



\subsection*{Theoretical scaling relation of the spiraling frequency with MT length and motor density}
The frequency of cytoskeletal spirals scaling behavior in gliding assay was theoretically predicted by balancing frictional force density of filament and propulsive force density of an 
implicit, {\it tangential} force generator \cite{bourdieu1995} to scale with force density as $\nu \sim f^{4/3}$. A similar scaling relation has also been devised and numerically demonstrated for self-propelled filaments \cite{fily2020buckling}. 
The velocity of filament driven by implicit force generators is directly proportional to the magnitude of the propulsion force. However, unlike implicit force generators, experimental gliding velocities of the filament driven by kinesin and dynein were reported to remain unaffected both by motor density and length of gliding filament \cite{Jain2019,monzon2019activation,howard1989movement}. In our simulations translational constraints, gliding velocity is invariant to both motor density and MT length (Figure \ref{fig:constant-gliding-velocity}) and thus is more realistic to gliding assay. Thus, the previously proposed scaling argument might not agree with the experimental data. Therefore, we propose an alternative scaling relationship assuming the spiraling phenomenon as uniform circular motion and thereby using the relation between angular frequency $\omega$ and linear velocity V, where $ \omega = V/R \nonumber$ from kinematics of circular motion. The spiraling frequency then become:
\begin{equation}
     |\omega| \sim \nu \sim \frac{|V_{tip}|}{R}
     \label{eq:freqgeom}
\end{equation}
Where $\nu$ is the frequency, and $V_{tip}$ is the linear velocity of the spiraling tip. The absolute value of the velocity, $|V_{tip}|$, is used to ensure physical consistency, since frequency cannot be negative, by definition. Without loss of generality, we assume that the tip velocity depends on both motor density and filament length. 
Substituting the previously derived scaling relation for the spiraling radius (Equation \ref{eqn:new-scaling-argument}), we obtain:
\begin{equation}
    \nu \sim  f^{1/3} |V_{tip}|(L,\rho_{m})/(k_{B}T\ell_{p})^{1/3}  
    \label{eqn:frequency_scaling_equation}
\end{equation}
This relation 
can be further simplified for the system where the {\it spiraling tip velocity of the filament is independent of both motor density and filament length}, yielding:
\begin{equation}
\nu \sim  |V_{tip}| \left(\frac{f}{k_{B}T\ell_{p}}\right)^{1/3} \implies  \nu \sim f^{1/3}
    \label{eqn:frequency_scaling_equation_simplfied}
\end{equation}

In the limiting case, where the velocity of the spiraling tip is linearly proportional to the force density,$|V_{tip}| \sim f$, we recover the scaling argument proposed in the existing literature:
\begin{equation}
    \nu \sim f\left(\frac{f}{k_BT\ell_p}\right)^{1/3} \implies 
    \nu\sim f^{4/3}
\end{equation}
We proceed to test the predicted relation obtained from scaling in terms of the effect of MT length and motor density on the spiraling frequency of the spirals using data described in the preceding section of MT tip position tracked in simulation and experiment. 

\subsection*{Frequency of filament spiraling scales with motor density deviates from a $4/3$ exponent in experiment and simulation} 
We computed the spiraling frequency of microtubule (MT) simulated with both constant and variable persistence lengths and corroborated these findings with experimental data. To derive a theoretical argument for scaling of spiraling frequency as a function of force density, we used (Equation \ref{eqn:frequency_scaling_equation}) and defined the scaled frequency as $\chi \equiv \nu(k_BT\ell_p)^{1/3}$
, which yields  the following relation:
\begin{equation}
    \chi \sim |V_{tip}|f^{1/3}
\end{equation}
We scaled the frequency as
$\chi_c \equiv \nu (\ell_p ^{c}k_B T)^{1/3}$ and $\chi_v \equiv \nu (\ell_p^v k_B T)^{1/3}$ for constant and variable persistence length simulations, respectively, using raw data (Figures \ref{fig:raw-frequency}A-C). We then used a power-law fit, $\chi_{c/v} \sim f^{\alpha}$, to estimate the scaling exponent.  Scaling analysis shows, rescaled frequency scales with force density with the exponents  $\alpha = 0.388$ and $\alpha=0.378$ for constant and variable persistence length simulations, respectively (Figure \ref{fig:freq_scaling}A,B). 


We observe a deviation from the theoretically predicted scaling exponent of $1/3$ in both constant and variable persistence length simulations. However, this deviation decreases as the stiffness of the pivot increases from 1000 to 10,000~pN/$\mu$m, with the scaling exponent approaching the theoretical value of $1/3$ (Figure \ref{fig:effect-of-trap-stiffness-frequency}). In the rescaled experimental data, analyzed using both constant and variable persistence lengths, we find a scaling exponent of $0.083$ and $0.369$, respectively (Figure~\ref{fig:freq_scaling}C). While this value deviates from the theoretical prediction of $1/3$, it is closer to the value of $1/3$ than to the previously proposed value of $4/3$. Based on the simulation with higher values trap stiffness, the deviation in experimental results likely arises from the finite stiffness of the pivot in the experimental setup.




\subsection*{Observed frequency scaling with MT length consistent with variable persistence length model}
\begin{figure}
\includegraphics[width=1\linewidth]{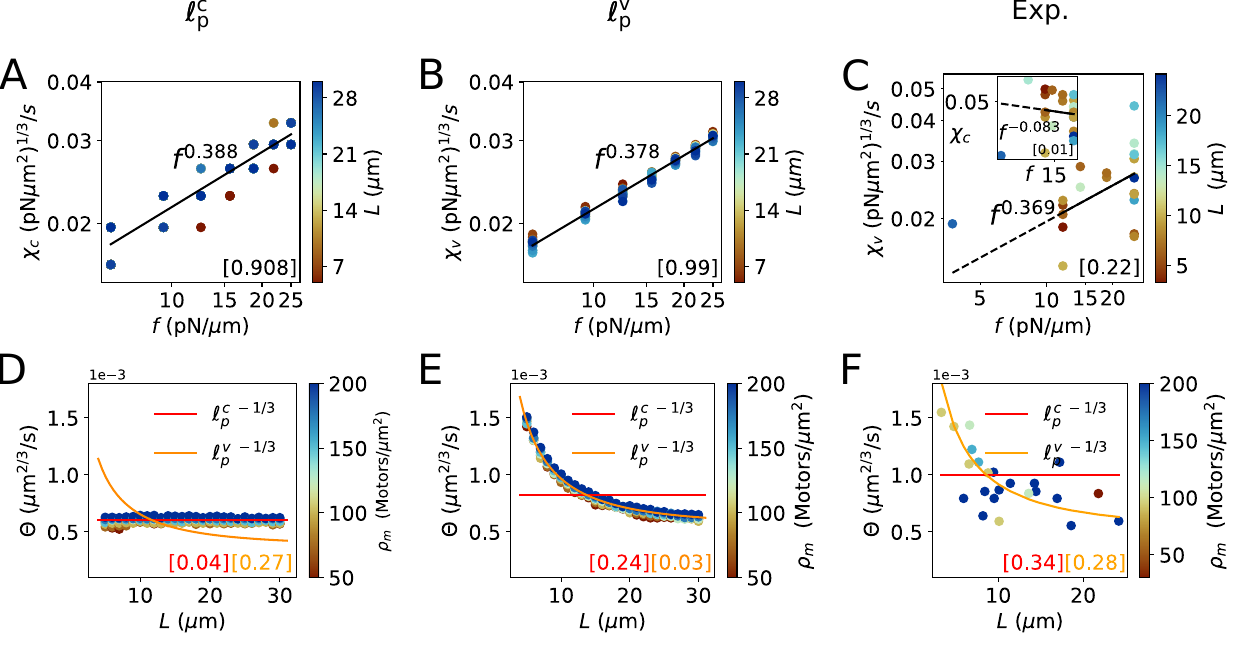}
\caption{\textbf{Scaling of spiraling frequency as a function of motor density and MT length} 
(A-C) Rescaled frequency of spiraling tip position, defined as either $\chi_c \equiv \nu (\ell_p^{c} k_B T)^{1/3}$ (assuming constant persistence length) or $\chi_p \equiv \nu (\ell_p^{v} k_B T)^{1/3}$ (assuming length-dependent persistence length), plotted as a function of force density ($f$), with microtubule (MT) length color-coded. (A, B) Simulations performed using (A) constant persistence length ($\ell_p^c$) or (B) variable persistence length ($\ell_p^v$). (C) Rescaled experimental data using variable persistence length (inset) shows rescaling using constant persistence length. A power law of the form $\chi_{c/v} \sim f^{\alpha}$ is fitted to the data (solid black line, ---), with the corresponding $R^2$ value displayed in the lower right corner. (D–F) Rescaled frequency $\Theta \equiv \nu (k_B T / f)^{1/3}$ of spiraling MTs is plotted as a function of MT length, with motor density color-coded. Scaling relations of the form $\Theta \sim \ell_{p}^{c~-1/3}$ and $\Theta \sim \ell_{p}^{v~-1/3}$ are fitted to the data and depicted using red and orange color lines, respectively, with the RMSRE (see method section) indicated in the lower right corner. Simulations were performed over MT lengths ranging from 5 to 30 $\mu$m and motor densities $\rho_m$ from 50 to 200 motors/$\mu$m$^{2}$, with $\ell_{p}^{c}$ set to 5 mm for constant persistence length (A,D) and $\ell_{p}^{v}(L) = 6.3/(1 + 441/L^2)$ mm for variable persistence length (B,E). Simulated data is represented by a mean of 20 repetitions.}

\label{fig:freq_scaling}
\end{figure}

The scaling of spiraling frequency as a function of filament length has not been previously reported in the literature. To understand the scaling behavior of both simulated and experimental data (Figure \ref{fig:raw-frequency}D-F), we begin our analysis by deriving the expected scaling relation using a theoretical scaling argument (Equation \ref{eqn:frequency_scaling_equation}). By rearranging the terms and defining a rescaled frequency as $\Theta \equiv \nu (k_BT/f)^{1/3}$, we arrive at the following scaling relation: 
\begin{equation} 
\Theta \sim |V_{tip}|\ell_p^{-1/3} 
\end{equation}
While the scaled frequency ($\Theta$) is not explicitly a function of filament length, it retains an implicit dependence via the persistence length $\ell_p$. Therefore, we expect the rescaled frequency to scale as a function of length in a manner proportional to $\ell_p^{-1/3}$. We tested the theoretical scaling argument against simulated data using constant and variable persistence length. In our analysis, we found that for simulations with constant persistence length, the scaling behavior was best described by $\Theta \sim \ell_p^{c~-1/3}$, as indicated by a lower root mean squared relative error (RMSRE: 0.04 $<$ 0.27; Figure \ref{fig:freq_scaling}D). Similarly, simulations with variable persistence length were better explained by the relation $\Theta \sim \ell_p^{v~-1/3}$ (RMSRE: 0.24 $<$ 0.03; Figure \ref{fig:freq_scaling}E). These results validate the theoretical scaling behavior as a function of filament length within the simulation data. We then examined the nature of scaling in the experimental data. When both models were fitted, the variable persistence length model provided a better explanation of the observed behavior (RMSRE: 0.34 $<$ 0.28; Figure \ref{fig:freq_scaling}F). This finding supports and reinforces our earlier conclusion that microtubule (MT) persistence length varies as a function of filament length.



\section*{Discussion}
In this study, we report for the first time spiraling patterns in a microtubule MT-motor system. These patterns were observed in a modified \textit{in vitro} gliding assay, enriched for spiraling using biotin–streptavidin chemistry, with yeast cytoplasmic dynein as the motor. These patterns appear to show a regularity and consistency of radius and frequency, which we attempt to explain using numerical simulation and theory. Our scaling analysis of the spiraling radius with motor density was found to result in an exponent that is consistent with previous work  \cite{bourdieu1995}.  The slight deviation of the experimentally observed scaling exponent ($\approx-1/2.7$) from the theoretical prediction ($-1/3$) is likely due to the finite stiffness of the pivot. In our simulations, the scaling exponent approached the theoretical prediction as the pivot stiffness was increased, supporting this interpretation.



Previous studies have examined similar motor-driven filament spiraling phenomena modeled using pinned polymers propelled by uniform tangential force density. Spiraling frequency was found to scale with force density to the power of $4/3$, supported by both simulations \cite{chelakkot2014flagellar,fily2020buckling} and scaling arguments \cite{bourdieu1995,fily2020buckling}. The divergence between existing literature and our theory is due to the oversimplification of motor dynamics in previous studies, which results in a deviation from biologically realistic phenomena. By using the kinematics of circular motion and radius scaling relation, as described in Equation~\ref{eqn:frequency_scaling_equation}, we arrive at the $1/3$ scaling. In arriving at this scaling relation, we assumed that the velocity of the spiraling free tip is invariant with respect to both motor density and MT length. This assumption is valid, as the variation in gliding velocities across the range of motor densities and MT lengths used in our simulations is small, and the velocities remain comparable to that of a single motor. In contrast, tangentially driven self-propelled filament models in a gliding assay result in velocities that are linearly proportional with force density, a behavior not typically observed in experimental gliding assay \cite{howard1989movement}. However, our scaling argument is general. When we assume the spiraling velocity to be linearly proportional to force density, we theoretically recover the $4/3$ scaling. The experimentally measured exponents ($\approx0.083 \text{ and } 0.369$) are closer to our proposed scaling argument ($1/3$) than to the value reported in previous literature ($4/3$).  Furthermore, the deviation from the theoretical scaling prediction ($1/3$) can be attributed to the finite stiffness of the pivot. This interpretation is supported by simulations, where increasing the pivot stiffness causes the exponent to approach the theoretical value. This result is also consistent with the scaling of the radius as a function of density, further reinforcing our interpretation.


The dependence of spiraling frequency and radius on MT length remains underexplored in the literature, primarily due to the common assumption of a constant persistence length. We theoretically predicted that the spiraling radius and frequency as function of length should scale with persistence length to the powers of $1/3$ and $-1/3$, respectively. After validating these scaling relationships, we examined experimental data to test the nature of MTs persistence length. We found that the scaling of both radius and frequency as functions of MT length fit better to the variable persistence length model, as indicated by a root mean squared relative error. These findings provide evidence that the persistence length of MTs is not constant, but rather varies with length.

A recent study reported beating-like oscillations in pinned filaments parameterized for actin \cite{khosravanizadeh2025dynamic}. Interestingly, our simulations of microtubules with dynein also showed similar transient wave-like reversals under certain limiting conditions (data not shown). However, within the length and density ranges relevant to our experiments and simulations, such transitions were not observed. This may be due to differences in parameters such as filament length, motor density, stall force, or linker stiffness. While our study focuses on exploring the general scaling laws, these preliminary observations suggest further work could help predict which kind of motors and under what circumstances such patterns might also be observed with microtubule-motor systems.

In this study, we validate generality of size scaling of active spirals in cytoskeletal filaments with force, based on experiments with pinned microtubules driven by dynein motors. We develop a novel model for the scaling of spiraling frequency with motor density based on the independence of transport velocity from motor density and test it experimentally. Further, we show radius and frequency scaling as function of length suggest a model of variable persistence length of microtubules best explains the data. These results could help better understand the buckling dynamics of biological filaments and improve our theoretical understanding.


\section*{Materials and methods}
\subsection*{Purification and labeling of microtubules and motor proteins}
\textit{Tubulin:} Goat brain lysate was subjected to activity cycling in presence of high molarity PIPES to purify active tubulin as mentioned previously \citep{CastoldiPOpov,Jain2019}. Rhodamine and biotin labeled tubulin were prepared by incubating 5(6)-Carboxytetramethylrhodamine N-succinimidyl ester(Molecular Probes, Eugene, OR, USA) and (+)-Biotin N-hydroxysuccinimide ester respectively with polymerized tubulin followed by a  depolymerization-polymerization cycle as described previously \citep{Yadav2024}. 

\noindent \textit{Dynein:} A fusion construct containing minimal yeast cytoplasmic dynein with a ZZ tag for purification, GST for dimerization and GFP for visualization (zz-GFP-GST-Dyn1$_{331}$) described previously\citep{reckpeterson2006} was expressed by growing a batch culture of yeast VY208 induced with 2\% (w/v) galactose (HiMedia, India). The protein was affinity purified from the cell lysate using IgG beads (GE healthcare, Sweden) and eluted using TEV protease treatment.

\noindent \textit{Anti-GFP Nanobody:} The Anti-GFP nanobody was used to specifically immobilize dynein on glass surface via the the GFP on the dynein tail. The  pGEX6P1-GFP-Nanobody construct was a gift from Kazuhisa Nakayama (Addgene plasmid 61838 \url{http://n2t.net/addgene:61838}.
RRID:Addgene\_61838). The nanobody was expressed and purified as mentioned previously \citep{Katoh2015,Yadav2024}.

\subsection*{MT assembly and spiral assay}

\textit{MT assembly:} Plus end biotin labeled MTs were prepared by first polymerizing the minus end with 22 $\mu${M} unlabeled tubulin + 8 $\mu${M} rhodamine labeled tubulin in BRB80 and 10$\text{\%}$ glycerol at 37$^{\circ}$C followed by continuing the polymerization of plus end by 18.33 $\mu$M biotin tubulin + 6.67 $\mu$M rhodamine tubulin for 25 minutes. Free monomers were removed by centrifuging at 1,35,300 g in TLA 100.3 rotor (Beckman Coulter, CA, USA) and MT pellet re-suspended in BRB-80 and 20 $\mu$M Taxol and used immediately. For gliding assay MTs without the biotin tubulin were prepared as mentioned above but skipping the second polymerization step.  

\noindent \textit{MT spiral assays:} The MT spirals were reconstituted in a set up similar to the MT beating assays by dynein described in detail and described briefly here. The key difference in both the methods being presence of shorter biotin MT plus ends. This was achieved by reducing the incubation time MT plus end polymerization by biotin-tubulin. The assay was reconstituted in a double back tape based chamber. The chamber was coated by a mixture of Anti-GFP nanobody and streptavidin (8 minutes) followed by passivation with 1mg/ml casein (8 minutes). Varying amount of dynein (1.05-3 $\mu$g) was introduced in the chamber to get various surface densities of motors (30-95  motors/$\mu$m$^2$). The plus end biotin MTs were introduced in the chamber and allowed to land. MT motility and spiraling was recorded after addition of the Motility buffer (30 mM HEPES (HiMedia, India), 2 mM Mg-Acetate (Amresco, OH, USA), 50 mM K-Acetate (Fisher Scientific, India), 4 mM ATP, Antifade mix [0.005 mg Glucose Oxidase, 0.0015 mg Catalase, 7.2 mM Glucose in 100 $\mu$l 10x Phosphate buffered saline, PBS (SRL Chemicals, Mumbai, India)].  

\noindent \textit{High density spirals based on surface defects in gliding assays:} For densities higher than 100 motors/$\mu$m$^2$ MT pinning was enriched via surface defects instead of biotin streptavidin strategy. To increase the pinning of MT due to surface defects, gliding assays with reduced timing for casein passivation was used. The gliding assays were performed in chambers as described above. The chamber was incubated with Anti-GFP Nanobody followed a brief passivation by casein (1-2 minutes). After the attachment of dynein, MTs were allowed to land. Motility was recorded after addition of Motility buffer. MT spirals with dynein densities of 100 - 208  motors/$\mu$m$^2$ were reconstituted using this method. Motor densities were estimated using an EGFP based calibration method as described in detail in \citep{Jain2019,Yadav2024}. All reagents unless otherwise stated, were from Sigma-Aldrich, MO, USA.

\subsection*{Microscopy and image-analysis}
The spiraling filaments were imaged in TRITC filter using 60x (NA=1.45) Oil immersion lens on a Nikon Ti-E inverted microscope (Nikon, Tokyo, Japan). The temperature was maintained at 37$^\circ$C using temperature control system (Okolab, Pozzuoli, Italy). The filaments were imaged every 10 seconds for 10 minutes using an Andor Clara2 CCD camera (Andor Technology, Belfast, UK). The images were denoised by applying median filter followed by background subtraction using FIJI \citep{schneider_rasband_eliceiri_2012}. 

The MT spiral diameter was measured manually using the measure tool of FIJI. For measuring the spiral diameter a transverse line was drawn covering the widest distance in the spiral. The radius was calculated from the diameter. For frequency analysis the free MT tip (Minus end) was tracked in time using MTrackJ plug in of FIJI (Figure \ref{fig:freq_track}A). From the position vs time data of the spirals the X and Y position with time was used to calculate the frequency of oscillations. The raw data were smoothed using cubic interpolation (Figure \ref{fig:freq_track}B) via the {\it interp1d} function from the {\it scipy.interpolate} package. Tracks were manually pruned to remove kinks and to ensure similar amplitude and wavelength. The interpolated data were then subjected to FFT analysis using the {\it scipy.fft} package to identify the dominant frequency (Figure \ref{fig:freq_track}C). Tracks exhibiting a divergence in dominant frequency greater than 0.01 Hz in the FFT spectrum were excluded from the scaling analysis.  

\subsection*{Estimating spiraling radius and linear force density}
\textit{Spiraling radius estimation:} A spiraling microtubule can be approximated as a limiting circle. For obtaining the spiraling radius, we use the Euclidean distance of contour points from the pinned end to estimate the spiraling radius of the pinned microtubule. To acquire robust statistics, we performed both spatial and temporal averaging of the distance from the pinned end to points between the $50^{th}$ and $100^{th}$ percentiles of the contour over steady-state spiraling configurations (Figure \ref{fig:averagin-radius-method}).\\

\noindent \textit{Conversion between motor density and linear force density:} In order to convert area density of motors ($\rho_m$) to linear force density, we apply some simplifications consistent with previous work \cite{bourdieu1995}. These simplifications involve estimates and assumptions about the potential number of motors interacting with an MT, the duration of force exerted by the motor and contribution of individual motors. The {\it potential number of motors interacting} with an MT is proportional to the dimensions of MT, i.e. $N_{mot} \propto L \cdot w$, where $L$ and $w$ are length and width of a microtubule. For a given density of molecular motor, $\rho_{m}$, the number of motors that can possibly interact are:

\begin{equation}
N_{mot} = L \cdot w  \cdot \rho_{m}
\label{numMot}
\end{equation}

based on the band model of cytoskeletal filament gliding assays \cite{Uyeda:1990aa}. The total force experienced by an MT, $F$, is proportional to number of motors interacting, for the duration of the interaction resulting in:

\begin{equation}
F = N_{mot} \ F_{o}  \cdot r
\label{totalMotorforce}
\end{equation}

where $r$ is the duty ratio and $F_{o}$ force contribution of each motor. The area density of motors are measured in experiments and are an input in simulations. Rearranging the previous expression (Equation \ref{totalMotorforce}), we obtain a relation for the linear force density ($f$) that depends on the motor density ($\rho_{m}$) as: 
\begin{equation}
    f = w \cdot r \cdot \rho_{m} \cdot F_{o}
    \label{eq:linforcedens}
\end{equation}
This approximation explicitly assumes that motor proteins act independently and represents the maximal force density that can be produced by the motors. Although it is a simplification, it can be used to estimate the force density applied by discrete force generators in the continuum limit for estimating scaling relations.

\noindent \textit{Goodness of fit of model:} For linear models, the \textit{coefficient of determination} ($R^2$) is an indicator of goodness of fit:
\[
R^2 = 1 - \frac{\sum (y - \hat{y})^2}{\sum (y - \bar{y})^2}
\]
However, for non-linear models, $R^2$ can be misleading. Instead, we use the \textit{root mean square of  relative error} (RMSRE), defined as:
\[
\text{RMSRE} = \sqrt{\frac{1}{N}\sum \left( \frac{y - \hat{y}}{y} \right)^2}
\]
This metric captures the relative squared deviation between the predicted and observed values, providing metric for comparing non-linear models.

\section*{Simulations}
Simulations were performed on Intel's Xeon Cascade Lake 2.9 GHz processors on a single processor at a time. The system RAM per node used is 192 GB using Param Bramha cluster \url{https://nsmindia.in/node/157\#1}. Typical simulations were run for 1200 seconds (20 min) which required between 1 to 11 hours for the two representative parameter sets: (i) $L$=10 $\mu$m, $\rho_{m}$=75 motors/$\mu$m$^{2}$ and (ii) $L$ = 29 $\mu$m, $\rho_{m}$=200 motors/$\mu$m$^{2}$, i.e. simulation time increases with motor density and filament length. 

\section*{Acknowledgments}
AS is supported by a fellowship from the Dept. of Biotechnology, Govt. of India, (DBT/2021-22/IISER-P/1851), SAY is supported by a fellowship from the CSIR, Govt. of India (09/936(0261)/2019-EMR-1). Computational resources provided by PARAM Brahma facility under the National Supercomputing Mission, Government of India at the IISER, Pune are gratefully acknowledged. We are grateful to Raghunath Chelakkot for discussions.

\section*{Author Contributions}

\section*{Data availability statement}
All the data has been made available with this manuscript. Additional raw data is available on request with the authors. Representative simulation parameter files and analysis scripts used for analyzing the radius and frequency dependence, along with scripts used for plotting the data, are available in a GitHub repository here: \url{https://github.com/CyCelsLab/spiraling}. 

\bibliographystyle{unsrtnat}
\bibliography{dynein-spiral}
\clearpage
\newpage
\section*{Tables}

\begin{table*}[ht!]
	\begin{center}
		\begin{tabular}{ p{2cm} p{5cm} p{4cm} p{4cm} }
			\textbf{Symbol} & \textbf{Description} & \textbf{Value} & \textbf{Reference} \\ \hline 
			\\
			{\it Microtubules:} & & & \\
			$\ell_{p}^{c}$     &  Constant persistence length  & 5$\times 10^3$ $\mu$m &  \citep{gittes1993flexural} \\
            $\ell_{p}^{v}(L)$    &  Variable persistence length  & $6300\left({1+ \frac{441}{L^{2}}}\right)^{-1}$ $\mu$m   &  \citep{pampaloni2006thermal}, Equation \ref{eq:pampaloniK} \\
			$L$     &  MT length              & 5 to 30 $\mu$m   & This study\\
			\\
   			{\it Dynein:} & & & \\
			${V}_{0}$     &  Single molecule motor velocity     & 0.10 $\mu$m/s     &  \citep{reckpeterson2006} \\
			$r_{a}$     &  Attachment rate                & 5  s$^{-1}$       &  \citep{Leduc2004} \\
			$d_{a}$     &  Attachment range               & 0.02 $\mu$m       &  Estimated from kinesin-binding range \citep{Hancock:1999vk} and molecular size of dynein. \\
			$r'_{d}$        & Load free detachment rate & 0.04 s$^{-1}$ & \cite{reckpeterson2006,Rao:2019aa} \\
			$F_{d}$     &  Detachment force               & 3 pN              &  \citep{Nicholas:2015wm}  \\
			$F_{s}$     &  Stall force                    & 5 pN              &  \citep{gennerich2007} \\ 
			$k_{m}$     &  Linker stiffness           & 100 pN/$\mu$m     & \citep{Lindemann:2003vn,Burgess:2003ut,Kamiya:2016uz} \\
			$\rho_{m}$ & Motor density & 50 to 200 motors/$\mu$m$^{2}$ & This study \\ 
				\\
            {\it Pivot:} & & & \\
           
            $k_m^p$ &Linker stiffness&10$^3$ to 10$^4$ pN/$\mu$m & This study \\
            \hline
		\end{tabular}
		\caption{{\bf Microtubule and motor parameters used in simulations.} The values for mechanochemical properties of the MTs, motors and the simulation system are taken from literature where available.}
		\label{tab:simparam}
	\end{center}	
\end{table*}

\clearpage
\newpage


\section*{Figures}
\clearpage
\newpage

\clearpage
\newpage





\clearpage
\newpage

\clearpage
\newpage
\subsubsection*{Supplementary Figures}
\renewcommand{\figurename}{Figure}
\renewcommand{\thefigure}{S\arabic{figure}} 
\renewcommand{\theHfigure}{S\arabic{figure}}   
\setcounter{figure}{0}

\begin{figure}[ht!]
    \centering
    \includegraphics[width=1\linewidth]{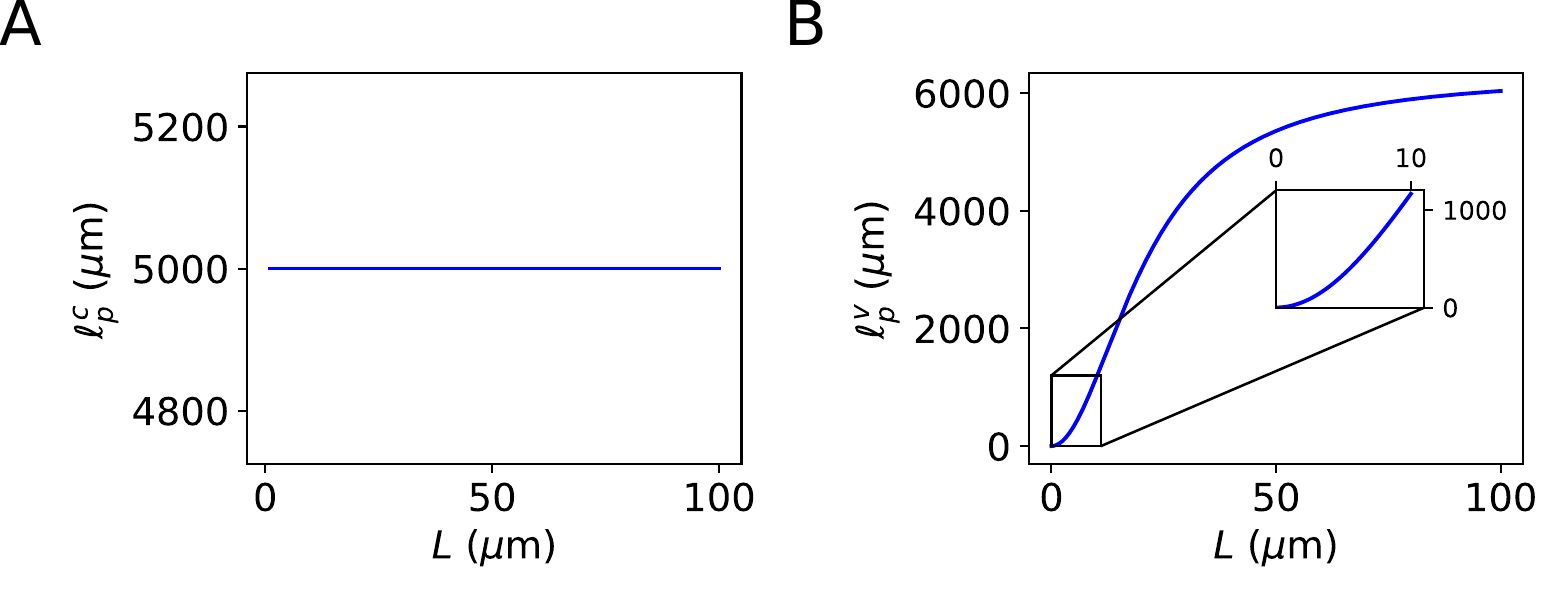}
    \caption{{\bf Comparing length-dependent and independent persistence lengths of MTs.} (A, B) The persistence length (y-axis)  as a function of MT length is plotted based on (A) the variable persistence length model (Equation \ref{eq:pampaloniK}) from work by Pampaloni et al. \cite{pampaloni2006thermal} with (inset) $\ell_p^{v}$ for physiological MT lengths of 1 to 10 $\mu$m and (B) constant persistence length based on the results of Gittes et al. \cite{gittes1993flexural}.}
    \label{fig:persistence-length-plots}
\end{figure}




\begin{figure}
    \centering
    \includegraphics[width=0.8\linewidth]{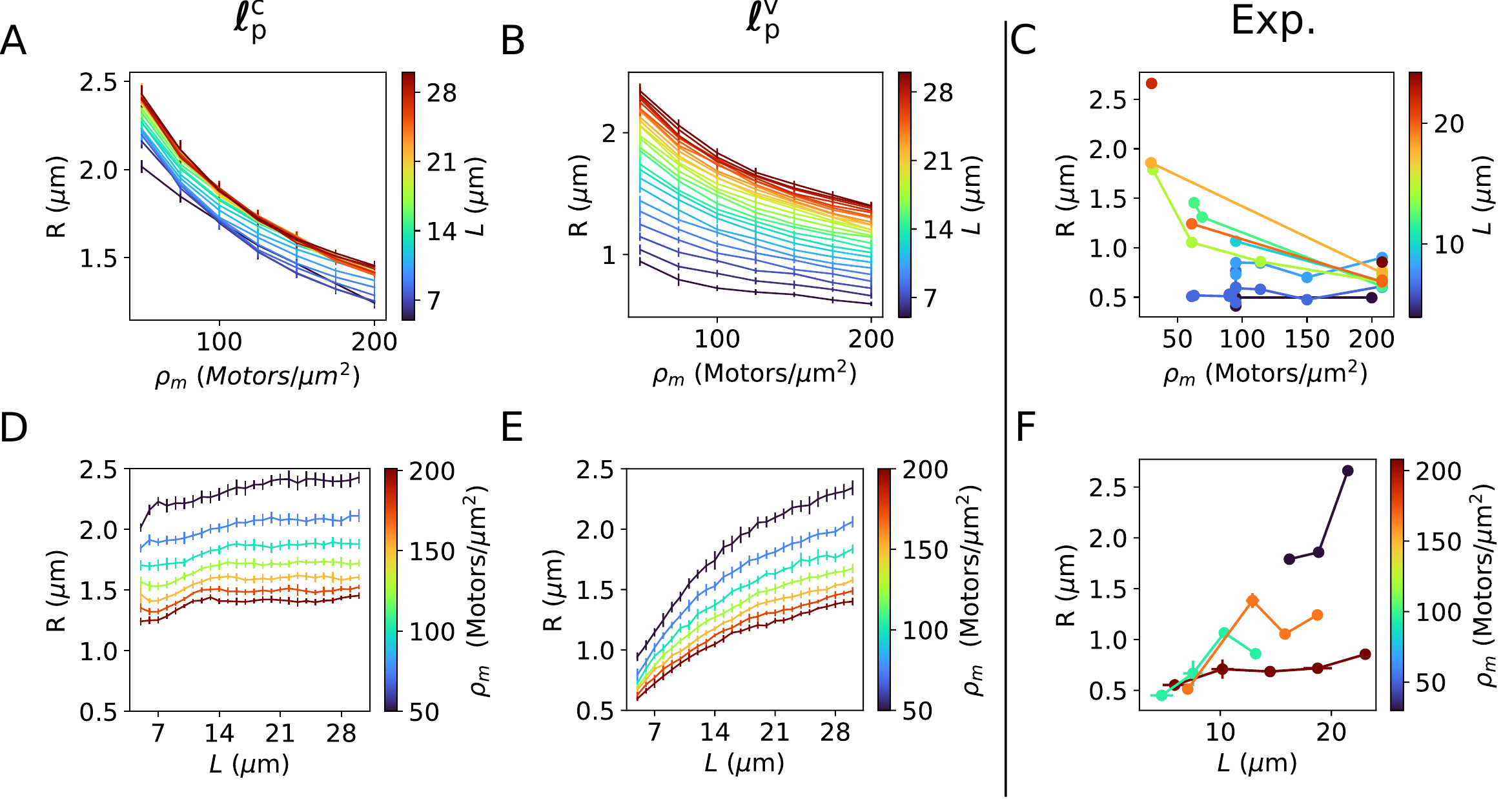}
\caption{\textbf{Radius analysis of spirals.} (A–C) Spiraling radius as a function of motor density with MT lengths (color-coded) plotted for (A) simulations with constant persistence length $(\ell_{p}^{c})$, (B) simulations with variable persistence length $(\ell_{p}^{v})$, and (C) experimental results. (D–F) Spiraling radius as a function of MT length, with motor densities (color-coded). (D) Simulations with constant $\ell_{p}^{c}$, (E) simulations with variable $\ell_{p}^{v}$, and (F) experimental results.}

    \label{fig:raw-radius}
\end{figure}

\clearpage
\newpage

\begin{figure}
    \centering
    \includegraphics[width=0.8\linewidth]{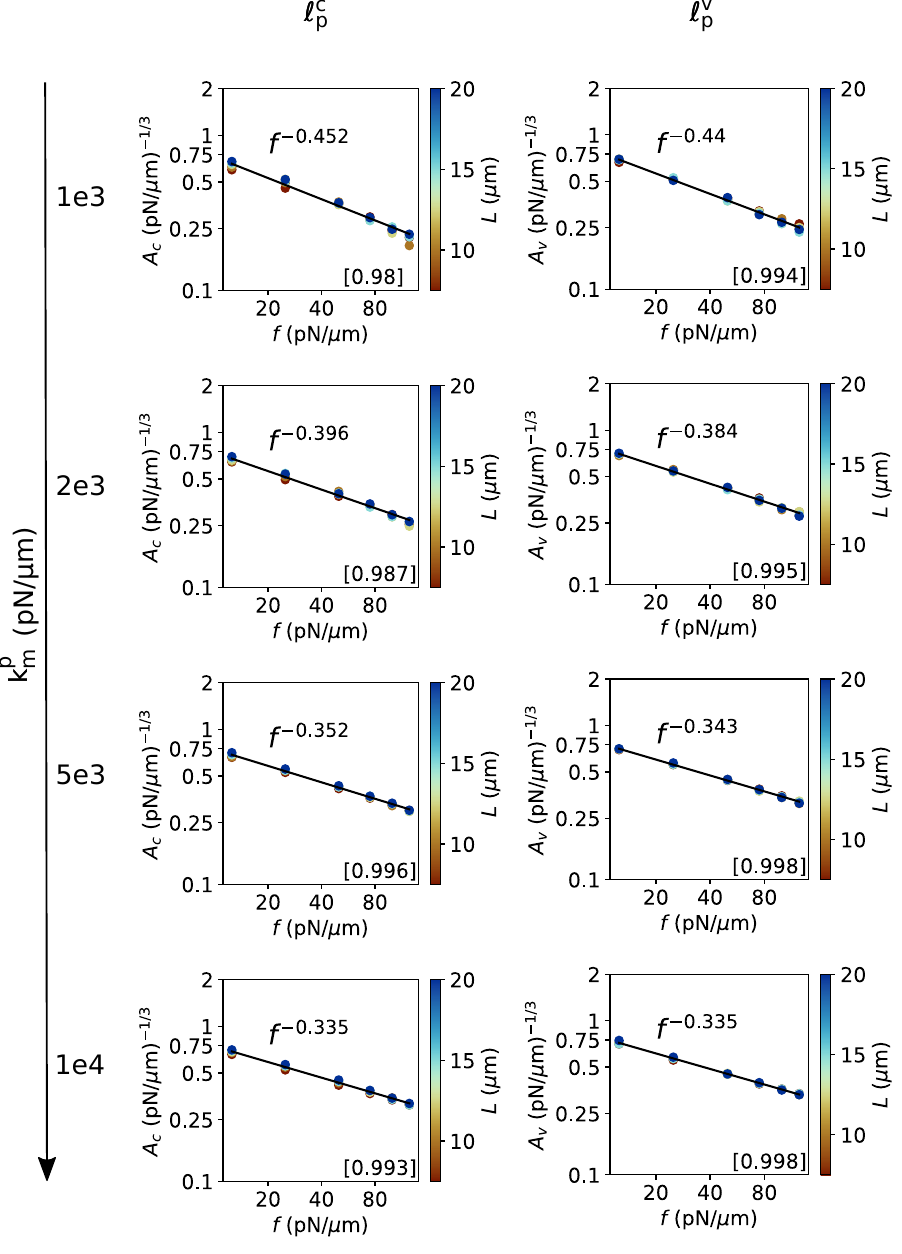}
    \caption{ \textbf{Effect of pivot Stiffness on spiraling radius scaling with force density.}  Simulations were performed for varying pivot stiffness values, with microtubule lengths $L = 7.5$, $10$, $12.5$, $15$, and $20\,\mu$m, and motor densities ranging from $100$ to $1000$ motors$/\mu$m$^{2}$. Scaling was analyzed via power-law fits of the form $A_{c/v} \sim f$. Scaling exponents obtained from the fits 
are indicated in the figure center legend with $R^2$ value depicted in the lower right corner. n=5}
    \label{fig:effect-of-trap-stiffness-radius}
\end{figure}


\clearpage
\newpage

\begin{figure}
    \centering
    \includegraphics[width=0.8\linewidth]{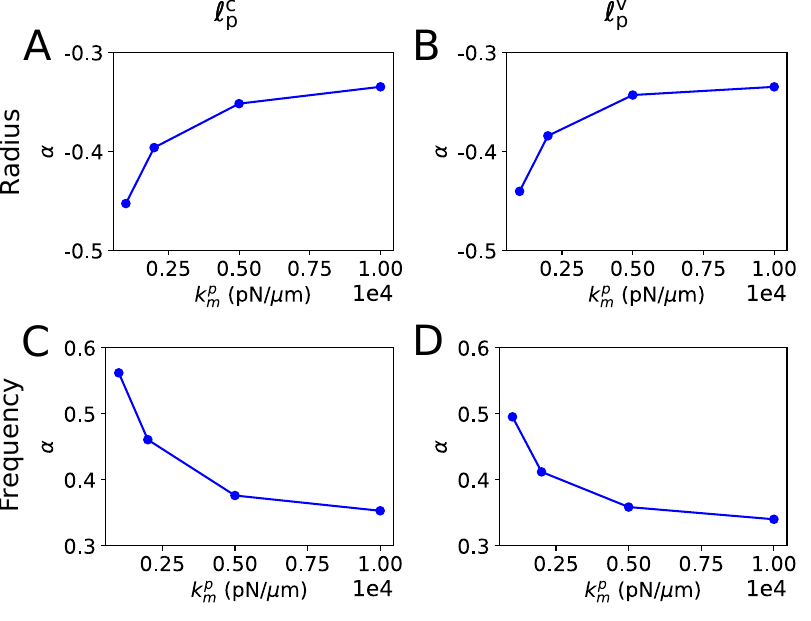}
    \caption{\textbf{Effect of pivot stiffness on the radius and frequency scaling exponents.} Simulations were performed with pivot stiffness values ranging from 10$^{3}$ to 10$^{4}$ pN$/\mu$m. We fitted power laws of the form $A \sim f^{\alpha}$, for radius scaling and $\chi \sim f^{\alpha}$, for frequency scaling, and plotted the corresponding exponents as a function of pivot stiffness.}
    \label{fig:scaling-stiffness-exponent}
\end{figure}

\clearpage
\newpage

\begin{figure}
    \centering
\includegraphics[width=0.8\linewidth]{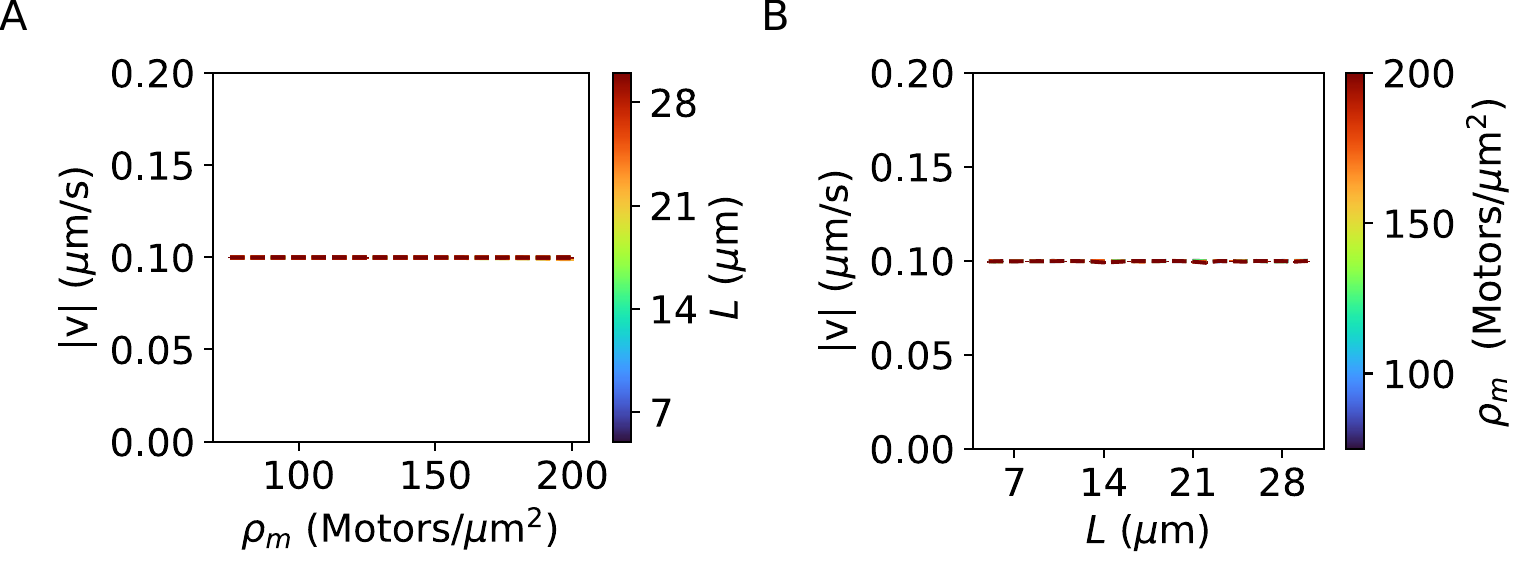}
    \caption{\textbf{Gliding velocity of microtubules (MTs) in \textit{in silico} gliding assay.} (A) Gliding velocity as a function of motor density (50--200 motors/$\mu$m$^{2}$), with different MT lengths ($L$) color coded. (B) Gliding velocity as a function of MT length (5--30 $\mu$m), with motor densities color coded.
}
    \label{fig:constant-gliding-velocity}
\end{figure}

\clearpage
\newpage

\begin{figure}
    \centering
    \includegraphics[width=0.8\linewidth]{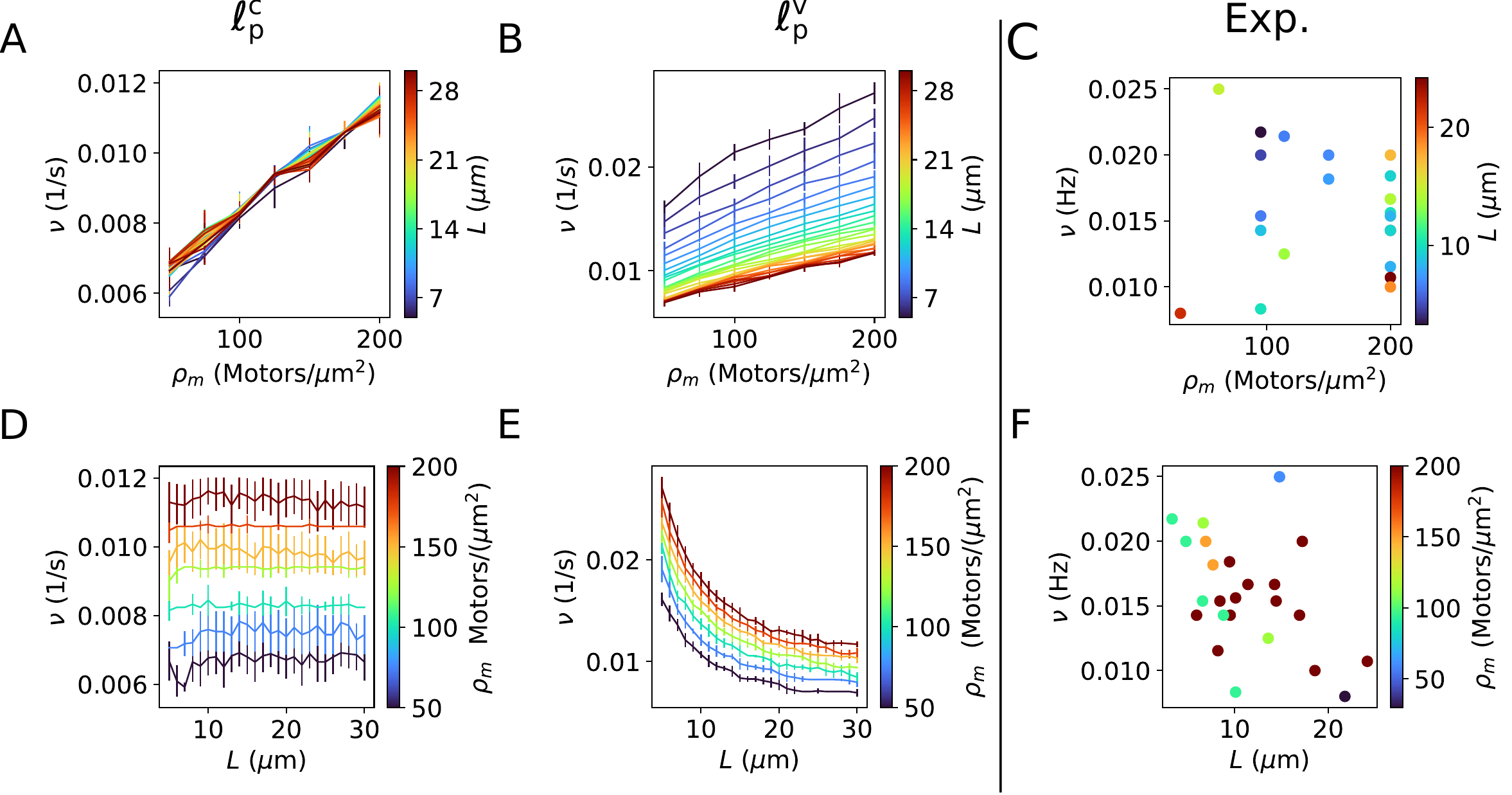}
    \caption{\textbf{Frequency analysis of spirals.} (A–C) Frequency of the MT free end plotted as a function of motor density, with MT lengths color-coded. (A) Simulations with constant persistence length $(\ell_{p}^{c})$, (B) simulations with variable persistence length $(\ell_{p}^{v})$, and (C) experimental results. (D–F) Frequency of the MT free end plotted as a function of MT length, with motor densities color-coded. (D) Simulations with constant $\ell_{p}^{c}$, (E) simulations with variable $\ell_{p}^{v}$, and (F) experimental results.}

    \label{fig:raw-frequency}
\end{figure}


\clearpage
\newpage

\begin{figure}
    \centering
    \includegraphics[width=0.8\linewidth]{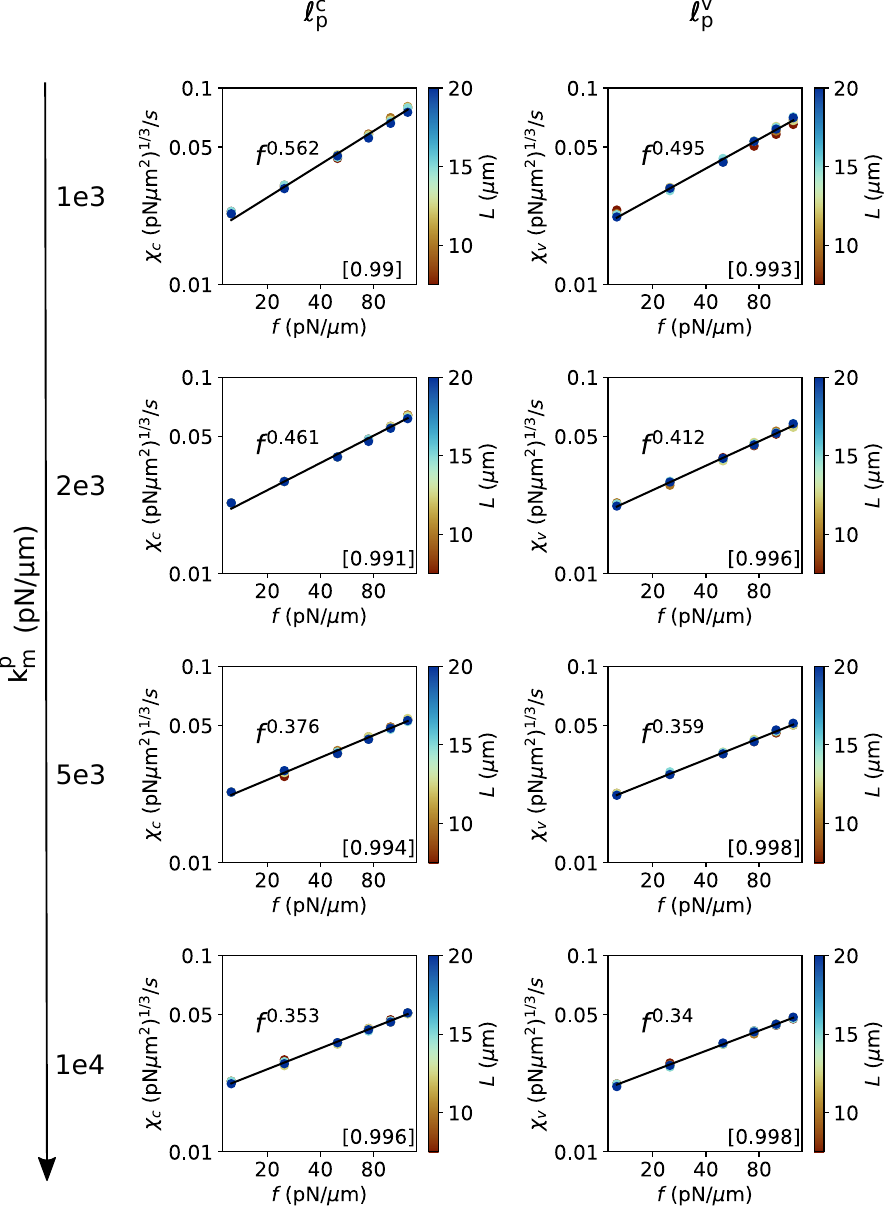}
\caption{\textbf{Effect of pivot stiffness on spiraling frequency scaling with force density.} Simulations were performed for varying pivot stiffness values, with microtubule lengths $L = 7.5$, $10$, $12.5$, $15$, and $20\,\mu$m, and motor densities ranging from $100$ to $1000$ motors$/\mu$m$^{2}$. Scaling was analyzed via power-law fits of the form $\chi_{c/v} \sim f$. Scaling exponents obtained from the fits are indicated in the figure center with $R^2$ value depicted in the lower right corner. n=5}
    \label{fig:effect-of-trap-stiffness-frequency}
\end{figure}

\begin{figure}
    \centering
    \includegraphics[width=0.8\linewidth]{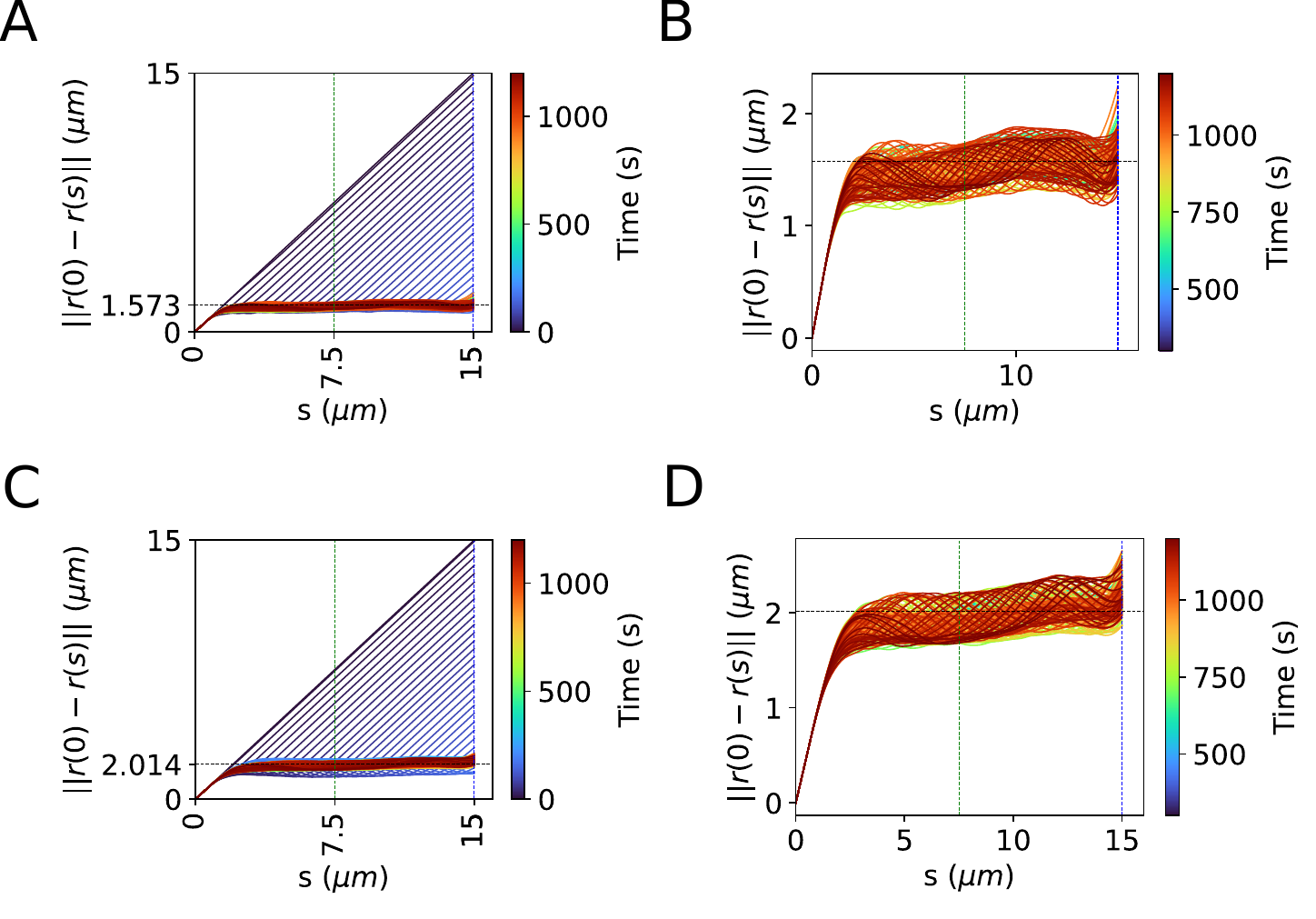}
    \caption{\textbf{Estimation of the spiraling radius of the pinned microtubule.}  We use the Euclidean distance of contour points from the pinned end to estimate the spiraling radius of the pinned microtubule. The dashed green and blue lines in the plots depict the $50^{th}$ and $100^{th}$ percentiles of the MT contour, respectively.  Figures (A $\&$ B) present data for a variable persistence length ($\ell_{p}^{c}$), while figures (C $\&$ D) show data for a constant persistence length ($\ell_{p}^{v}$). In figures (A) and (C), the Euclidean distance is plotted as a function of time, with a color code representing the initial conditions. In panels (B) and (D), the Euclidean distance is shown for the steady-state spiraling time. The length of the microtubule ($L$) is 15 $\mu$m, and Cytoplasmic dynein density ($\rho_{m}$) is 75 motors/$\mu$m$^{2}$.}
    \label{fig:averagin-radius-method}
\end{figure}

\clearpage
\newpage

\begin{figure}
    \centering
    \includegraphics[width=0.8\linewidth]{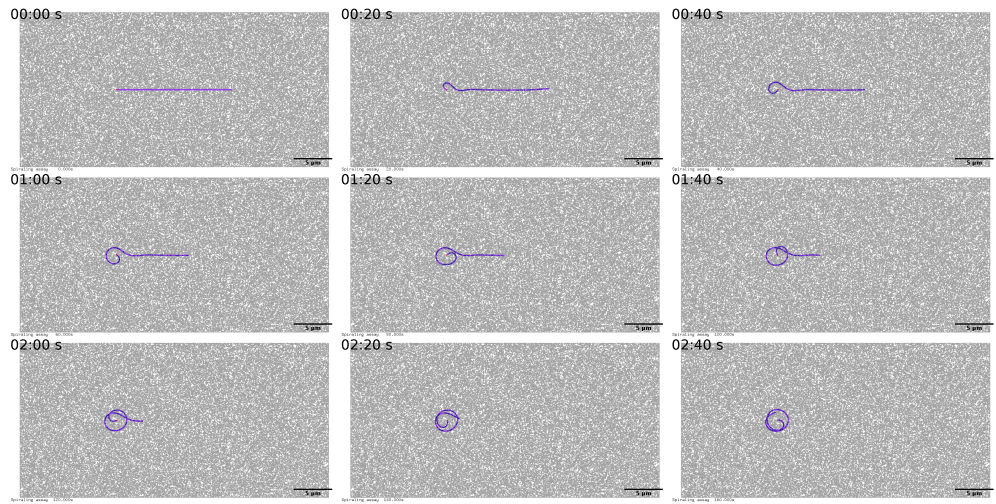}
    \caption{\textbf{Temporal evolution of MT in an \textit{in silico} spiraling assay.} Montage showing the evolution of MT from a straight configuration at the initial time point to a steady-state spiraling configuration. The images in the montage are separated by time intervals of 20 seconds. $L$ = 15 $\mu$m and cytoplasmic dynein density, $\rho_{m}$ = 100 motors/$\mu$m$^{2}$.}
    \label{fig:motange3x3-temporal-evolution}
\end{figure}

\clearpage
\newpage

\clearpage
\newpage

\subsubsection*{Supplementary Videos}
\renewcommand{\figurename}{{\bf Video}}
\renewcommand{\thefigure}{SV\arabic{figure}}  
\renewcommand{\theHfigure}{SV\arabic{figure}}  
\setcounter{figure}{0}

\begin{figure}[ht!]
    \centering
    \includegraphics[width=0.8\linewidth]{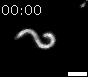}
    \caption{{\bf Single biotin-streptavidin pinned microtubule in a dynein gliding assay.} A representative rhodamine labeled MT showing spiral formation upon pinning of the leading tip due to stepping of dynein in presence of ATP in a dynein gliding assay. $L$= 6.8 $\mu$m, $\rho_m$ = 61/$\mu$m$^{2}$, Time: mm:ss, Scale bar: 4 $\mu$m. }
    \label{vid:spiral_expt}
\end{figure}

\clearpage
\newpage
\begin{figure}
   \centering
   \includegraphics[width=0.8\linewidth]{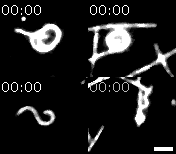}
   \caption{\textbf{Effect of motor density and microtubule length on spiral radius in experiments.} Representative rhodamine labeled MTs of $L$ = 19.2 $\mu$m (Top left), 6.8 $\mu$m (Bottom left), 16.9 $\mu$m (Top right) and 5.92 $\mu$m (Bottom right) forming spirals at dynein density of 61/$\mu$m$^{2}$ (Left) and 208/$\mu$m$^{2}$ (Right). The videos correspond to Figure \ref{fig:experimental-stepup} C. Scale bar = 2 $\mu$m. Time: mm:ss.}
\end{figure}

\clearpage
\newpage

\begin{figure}
    \centering
    \includegraphics[width=0.8\linewidth]{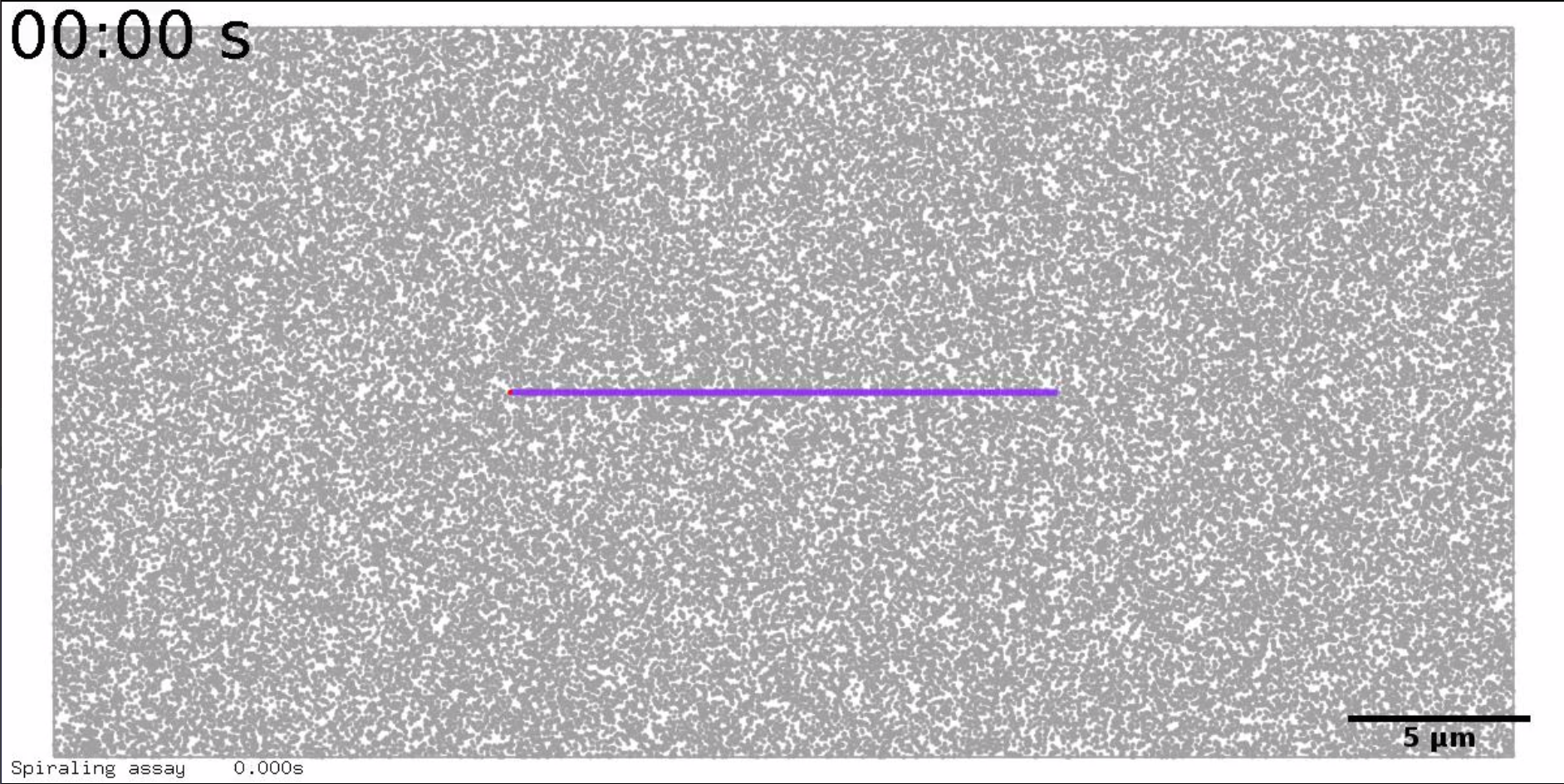}
    \caption{{\bf Simulating a single single plus end pinned microtubule in a dynein gliding assay.} The representative simulation consists of a MT (purple) of length 15 $\mu$m, pinned at the plus end by a pivot (red). Motors attached to the MT are shown in blue; the rest are depicted in gray. The simulation was performed at a motor density ($\rho_{m}$) of 100 motors/$\mu m^2$ for 1200 seconds. Scale bar: 5 $\mu$m; Time: mm:ss}
\end{figure}

\clearpage
\newpage
\begin{figure}
    \centering
    \includegraphics[width=0.8\linewidth]{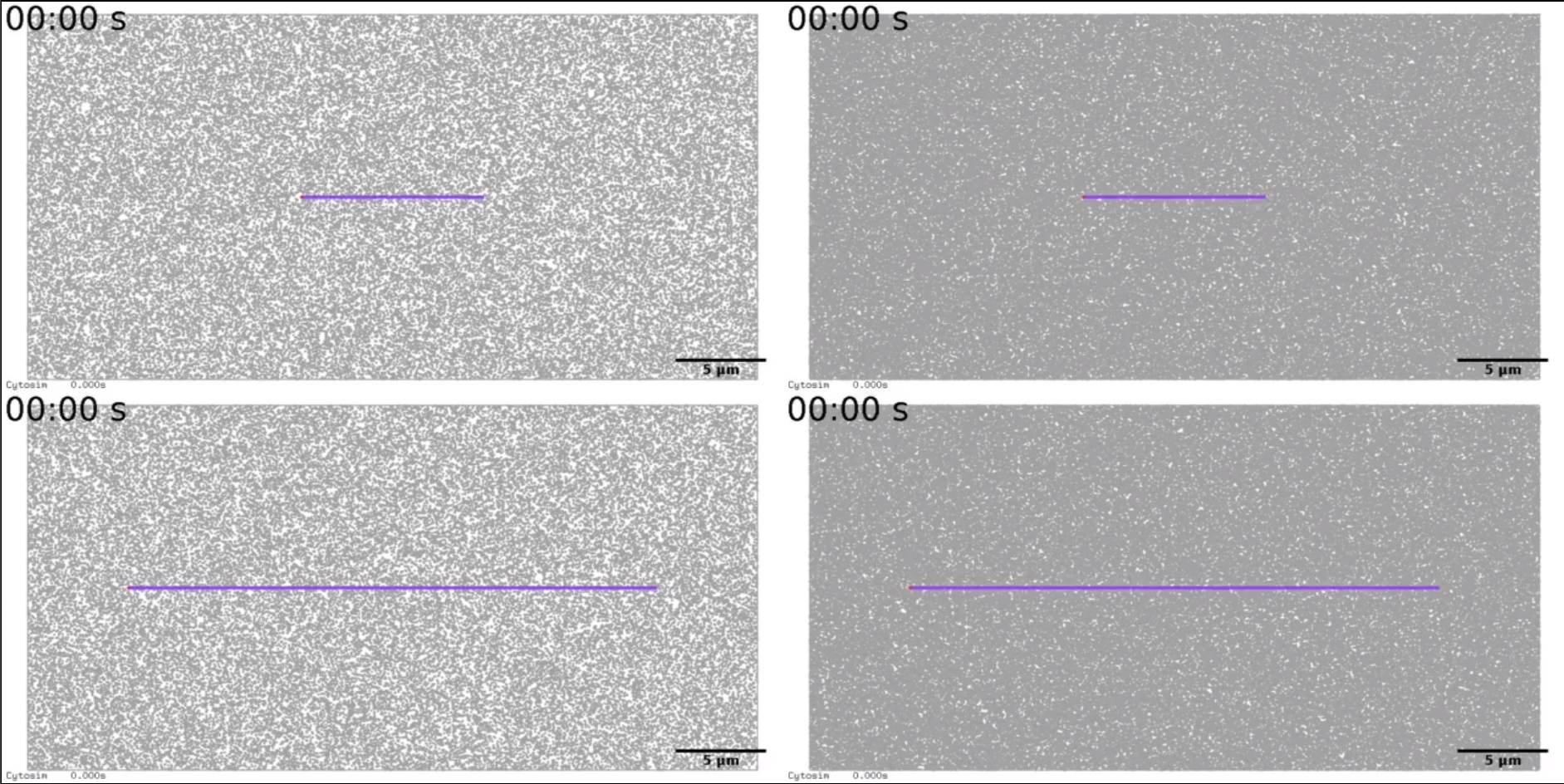}
    \caption{\textbf{Effect of motor density and microtubule length on spiraling patterns.} Representative simulations were conducted at motor densities of 75 motors/$\mu m^{2}$ (left) and 200 motors/$\mu m^{2}$ (right), with MT lengths of 10 $\mu$m (top) and 29 $\mu$m (bottom). Scale bar: 5 $\mu$m; Time: mm:ss}
    \label{fig:template-motange-video}
\end{figure}
\end{document}